\documentclass[extra,twocolumn]{gji}

\usepackage{amsmath,amsfonts,bm}
\usepackage{amssymb}
\usepackage{graphicx}
\usepackage{overpic}
\usepackage[colorlinks=true,citecolor=cyan]{hyperref}

\usepackage{xcolor}
\usepackage{ulem}

\author[F. Gerick, D. Jault, J. Noir and J. Vidal]
  {F. Gerick$^{1,2}$, D. Jault$^1$, J. Noir$^2$ and J. Vidal$^3$ \\
  $^1$ Univ. Grenoble Alpes, Univ. Savoie Mont Blanc, CNRS, IRD, IFSTTAR, ISTerre, 38000 Grenoble, France.\\
  $^2$ Institute of Geophysics, ETH Zurich, Zurich, Switzerland\\
  $^3$ Department of Applied Mathematics, School of Mathematics, University of Leeds, Leeds, LS2 9JT, United Kingdom
  }
\title{Pressure torque of torsional Alfv\'en modes acting on an ellipsoidal mantle}

\renewcommand{\vec}{\mathbf}

\newcommand{\B}{\vec{B}}
\newcommand{\vel}{\vec{u}}

\newcommand{\pos}{\vec{r}}
\newcommand{\posp}{\vec{r}_\perp}

\newcommand{\ez}{\vec{1}_z}
\newcommand{\rotax}{\vec{1}_\Omega}
\newcommand{\Dop}{\mathcal{D}}
\newcommand{\Qop}{\boldsymbol{\mathcal{Q}}}
\newcommand{\Le}{\mathrm{Le}}
\newcommand{\Lu}{\mathrm{Lu}}
\newcommand{\Pm}{\mathrm{Pm}}

\newcommand{\dV}{\mathrm{d}V}
\newcommand{\dA}{\mathrm{d}S}

\newcommand{\R}{\mathbb{R}}
\newcommand{\Cn}{\mathbb{C}}

\newcommand{\qgvel}{\vec{u}}
\newcommand{\torquec}{\vec{\Gamma}_\mathrm{c}}
\newcommand{\torquep}{\vec{\Gamma}_\mathrm{p}}
\newcommand{\torquepm}{\vec{\Gamma}_\mathrm{pm}}

\newcommand{\torqueb}{\vec{\Gamma}_\mathrm{L}}
\newcommand{\torquebgradb}{\vec{\Gamma}_\mathrm{b}}
\newcommand{\bcdot}{\boldsymbol{\cdot}}
\newcommand{\imwidth}{\linewidth}

\definecolor{forestgreen}{rgb}{0.13333333333333333, 0.5450980392156862, 0.13333333333333333}
\begin{document}
\maketitle

\begin{abstract}
We investigate the pressure torque between the fluid core and the solid mantle arising from magnetohydrodynamic modes in a rapidly rotating planetary core. A two-dimensional reduced model of the core fluid dynamics is developed to account for the non-spherical core-mantle boundary. 
The simplification of such a quasi-geostrophic model rests on the assumption of invariance of the equatorial components of the fluid velocity along the rotation axis.
We use this model to investigate and quantify the axial torques of linear modes, focusing on the torsional Alfv\'en modes (TM) in an ellipsoid. 
We verify that the periods of these modes do not depend on the rotation frequency. Furthermore, they possess angular momentum resulting in a net pressure torque acting on the mantle. This torque scales linearly with the equatorial ellipticity. We estimate that for the TM calculated here topographic coupling to the mantle is too weak to account for the variations in the Earth's length-of-day. 

\end{abstract}


\section{Introduction}

Decadal variations in the Earth's length-of-day (LOD) have long been associated with dynamics in the liquid outer core \citep{Munk1960,hide_free_1966,jault_westward_1988,gross_2015}.
More specifically, a pronounced variation on a period of roughly six years cannot be explained by atmospheric, oceanic and tidal forces, which are responsible for LOD variations on shorter time scales \citep{abarca_del_rio_interannual_2000, holme_characterization_2013}.
Torsional Alfv\'en modes (TM) in the outer core, first studied by \citet{braginsky_torsional_1970}, have been proposed later as the origin of the six year variation in the LOD \citep{gillet_fast_2010}.
In the sphere, these oscillations consist of differentially rotating nested geostrophic cylinders, stretching and shearing the magnetic field lines.
Recent advances in magnetic field observations and inverse modelling of the outer core flow at the core-mantle boundary (CMB) have revealed recurring TM with 4-year travel time through the Earth's outer core \citep{gillet_fast_2010,gillet_planetary_2015}.
\cite{gillet_excitation_2017} investigated the LOD variations that result from the TM propagation, assuming that the only stresses between the core and the mantle are electromagnetic. Relying on the study of \cite{Schaeffer2016}, they inferred constraints on the electrical conductivity of the lowermost mantle.
To account for the observed LOD variations, a conductance of the lowermost mantle of $3\times 10^7 - 10^8 \,  \mathrm{S}$ is needed \citep{gillet_excitation_2017}.
Another mechanism of coupling outer core dynamics to the solid mantle is through gravitational coupling between a deformed inner core and a non-spherical CMB \citep{buffett_gravitational_1996,buffett_mechanism_1996,  mound_detection_2006}. A phase lag between the deformations leads to a torque on the mantle. Even though recent advances in atmospheric and oceanic tide modelling have improved the isolation of gravitational signals from core dynamics, the measurements are still inconclusive \citep{davies_strength_2014, watkins_earths_2018}.

The third mechanism, investigated here, that may account for exchange of angular momentum between core and mantle is topographic coupling.
It has long been proposed that, for a non-spherical CMB, there could be a significant pressure torque exerted by flows in the outer core  \citep{hide_interaction_1969}.
The fluid pressure should scale as $\rho \Omega U R_0$, 
where $\rho$ is the core density, $\Omega$ the angular speed of the Earth's rotation, $U$ a typical horizontal velocity and $R_0$ the core radius. 
A typical amplitude of  $\mathcal{O}(10^3)$ Pa has been obtained from core surface velocity models, assuming a local balance of force (tangential geostrophy) at the core surface \citep{jault_core-mantle_1990}. 
These models have now been superseded by quasi-geostrophic (QG) models that rely on a global assumption, for which it is assumed that the equatorial components of the fluid velocity are invariant along the rotation axis, as observed at leading order in numerical simulations \citep[e.g.][]{gillet_rationale_2011, schaeffer_turbulent_2017}.
QG models have been shown to capture the fundamental features of rapidly rotating hydrodynamics by comparing with three-dimensional (3-D) numerical simulations \citep{guervilly2019turbulent,gastine2019pizza}.
Furthermore, QG models incorporating the magnetic field have been used to investigate spherical TM \citep{Canet2014,Labbe2015}.  In this framework, the surface pressure cannot be inferred from the velocity. 
In the most general case, the pressure is a 3-D quantity given by the Lagrange multiplier associated to incompressibility. 
For QG models we can introduce a Lagrange multiplier associated to incompressibility, but it is only a two-dimensional (2-D) function of the coordinates in the equatorial plane. Therefore, we cannot infer the 3-D pressure at the CMB from the velocity field only.

For an axisymmetric core, the axial pressure torque vanishes exactly for any flow. To investigate the influence of non-axisymmetric CMBs, the ellipsoidal geometry can be considered as a first step. 
From seismological observations a peak-to-peak amplitude of CMB topography of about 3 km has been inferred \citep{sze_core_2003, koper_constraints_2003}, corresponding to an equatorial ellipticity $\mathcal{O}(10^{-3})$.

Here, we derive a generic QG model that does not assume axisymmetry, which is then compared to a hybrid model using QG velocities and 3-D magnetic fields in the case of an ellipsoid.
We present the linear modes and their axial angular momentum, as well as the hydrodynamic pressure torque that the fluid exerts on the solid container.
Lastly, we discuss the possible implications of this study for Earth-like liquid cores.

\section{Problem setup}
\subsection{Magnetohydrodynamic equations}

We consider a fluid of homogeneous density $\rho$, uniform kinematic viscosity $\nu$ and magnetic diffusivity $\eta$, which is enclosed in a rigid container of volume $\mathcal{V}$ and boundary $\partial\mathcal{V}$.
The time evolution of the velocity field $\vel$ and the magnetic field $\B$ is given by the incompressible magnetohydrodynamics (MHD) equations. 
In the reference frame rotating with the angular velocity $\vec{\Omega}$, they read
\begin{subequations}
\label{eq:goveqsdim}
\begin{align}
    \begin{split}
        \label{eq:momeq}
    \frac{\partial \vel}{\partial t}+ ( \vel\boldsymbol{\cdot}\boldsymbol{\nabla} ) \, \vel =&    -2\, \vec{\Omega}\times\vel-\frac{1}{\rho}\nabla p + \nu \, \boldsymbol{\nabla}^2\vel  \\
    &+\frac{1}{\mu_0\rho} \, (\boldsymbol{\nabla}\times\B)\times\B ,
    \end{split}\\
     \frac{\partial \B}{\partial t} =& \boldsymbol{\nabla}\times\left(\vel\times\B\right)+\eta\, \boldsymbol{\nabla}^2\B,
\end{align}
\end{subequations}
with $p$ the reduced pressure and $\mu_0$ the magnetic permeability in vacuum.
MHD equations (\ref{eq:goveqsdim}) are completed by the solenoidal conditions $\boldsymbol{\nabla}\boldsymbol{\cdot} \B = \boldsymbol{\nabla} \boldsymbol{\cdot} \vel = 0$.
The characteristic length scale $R_0$ is determined by the container size, which
is taken as its mean radius.
In ellipsoids, $R_0$ is the geometric mean $R_0=(abc)^{1/3}$ of the three semi-major axes $[a,b,c]$. 
The angular velocity is given by $\vec{\Omega}=\Omega\rotax$ and the characteristic background magnetic field strength is $B_0$. We define the characteristic time $t_0 = R_0/u_A$, where $u_A=B_0/\sqrt{\rho \mu_0}$ is the characteristic Alfv\'en wave velocity. 
The characteristic pressure is then given by $\rho u_A^2$.
The dimensionless equations read
\begin{subequations}
\label{eq:goveqadimNL}
\begin{align}
    \begin{split}
        \label{eq:momeqnondim}
    \frac{\partial \vel}{\partial t} + (\vel\boldsymbol{\cdot}\boldsymbol{\nabla})\, \vel =& -\frac{2}{\Le} \, \rotax\times\vel  -\nabla p + \frac{\Pm}{\Lu}\, \boldsymbol{\nabla}^2\vel\\
    &+ (\boldsymbol{\nabla}\times\B)\times\B,
    \end{split}\\
    \frac{\partial \B}{\partial t} =& \boldsymbol{\nabla} \times\left(\vel\times\B\right)+\frac{1}{\Lu} \, \boldsymbol{\nabla}^2\B,
\end{align}
\end{subequations}
where we introduce the Lehnert number $\Le$ (measuring the strength of the Lorentz force relative to the Coriolis force), the Lundquist number $\Lu$ (comparing magnetic induction to magnetic diffusion), and the magnetic Prandtl number $\Pm$ (comparing kinematic viscosity to magnetic diffusion). 
They are given by 
\begin{equation}
    \Le=\frac{B_0}{\Omega R_0 \sqrt{\mu_0\rho}}, \quad \Lu = \frac{R_0B_0}{\eta\sqrt{\mu_0\rho}}, \quad \Pm = \frac{\nu}{\eta}.
\end{equation}
Typical values for the Earth's outer core, with radius $R_0\approx3478$ km, kinematic viscosity $\nu\approx 10^{-6}$ m$^2$s$^{-1}$ \citep{wijs_viscosity_1998}, mean radial magnetic field strength $B_0\approx 3$ mT \citep{gillet_fast_2010} and electrical conductivity $\sigma\approx 1.55\times 10^6$ Sm$^{-1}$ \citep{pozzo_thermal_2014}, are $\Le = \mathcal{O}(10^{-4})$, $\Lu = \mathcal{O}(10^5)$ and $\Pm = \mathcal{O}(10^{-6})$. 
The dynamics we will be considering operate on timescales shorter than magnetic diffusion and  viscous spin-up times. Hence, we will neglect viscous and Ohmic dissipations.
The governing equations are
\begin{subequations}
\label{eq:goveqadimideal}
\begin{align}
    \frac{\partial \vel}{\partial t} + (\vel\boldsymbol{\cdot}\boldsymbol{\nabla}) \, \vel =& - \frac{2}{\Le} \, \rotax\times\vel-\nabla p +(\boldsymbol{\nabla}\times\B)\times\B,\label{eq:momeqnondimND} \\
    \frac{\partial \B}{\partial t} =& \boldsymbol{\nabla} \times\left(\vel\times\B\right).\label{eq:indeqideal}
\end{align}
\end{subequations}

Equations (\ref{eq:goveqadimideal}) are supplemented with appropriate boundary conditions. 
In the diffusionless approximation, the velocity needs to satisfy only the non-penetration condition  $\vel\boldsymbol{\cdot}\vec{n}=0$ on $\partial V$.
If $\B\cdot\vec{n}=0$ at an initial time $t=0$, the normal component of the induction equation ensures that the normal component of $\B$ is zero at all later times \citep[see][]{backus1996foundations}.

\subsection{Torque balance}

The net torque balance of the system is given by
\begin{equation}
    \frac{\partial \vec{L}}{\partial t} +\torquec = \torquep + \torqueb,\label{eq:torquebalance}
\end{equation}
with the angular momentum $\vec{L}$, the hydrodynamic pressure torque $\torquep$, the Coriolis torque $\torquec$ and the Lorentz torque $\torqueb$ given by
\begin{subequations}
\begin{align}
    \vec{L} &= \int_\mathcal{V}\pos\times \vel\, \dV, \label{eq:angularmom}\\
    \torquep &= -\int_\mathcal{V}\pos\times  \nabla p\, \dV = -\int_{\partial\mathcal{V}} p\, (\pos\times\vec{n})\, \dA, \label{eq:hydropressuretorque}\\
    \torquec &= 2\int_\mathcal{V}\pos\times(\vec{\Omega}\times\vel)\, \dV,\label{eq:torquec}\\
    \torqueb &= \int_\mathcal{V}\pos\times(\left(\boldsymbol{\nabla}\times\B\right)\times\B)\, \dV.
\end{align}
\end{subequations}
We can further split up the Lorentz torque into magnetic pressure torque $\torquepm$ and a magnetic tension torque $\torquebgradb$ as
\begin{equation}
    \torqueb = \torquebgradb+\torquepm,
\end{equation}
with
\begin{subequations}
\begin{align}
    \torquepm &= -\frac{1}{2}\int_\mathcal{V} \pos\times\nabla\left(\B^2\right)\,\dV=-\frac{1}{2}\int_{\partial\mathcal{V}} \B^2(\pos\times\vec{n})\, \dA,\label{eq:torquepm}\\
    \torquebgradb &=\int_\mathcal{V} \pos\times((\B\boldsymbol{\cdot}\nabla) \, \B)\, \dV.\label{eq:torquebgradb}
\end{align}
\end{subequations}
For a perfectly conducting boundary (with  $\B\boldsymbol{\cdot}\vec{n}=0$ on $\partial\mathcal{V}$), $\torquebgradb$ vanishes exactly \citep[see equation 45 in][]{roberts_theory_2012} and only the magnetic pressure torque $\torquepm$ contributes to the torque balance \eqref{eq:torquebalance}.

The axial component of the Coriolis torque $\torquec$ also vanishes \citep[see equation 14.98 in][]{davidson_introduction_2016}. In the axial direction, the torque balance reduces to
\begin{equation}
    \frac{\partial L_z}{\partial t} = \Gamma_{\mathrm{p},z}+\Gamma_{\mathrm{pm},z}. \label{eq:remainingbalance}
\end{equation}
Hence, any changes of the axial angular momentum of the fluid can only result from the unbalance between the magnetic and hydrodynamic pressure torques.
For the sphere, the transformation of the volume integral into a surface integral shows that the pressure torques \eqref{eq:hydropressuretorque} and \eqref{eq:torquepm} vanish, so that no change in angular momentum is possible.

\subsection{Geostrophic motions and torsional Alfv\'en modes}
\label{subsec:pressuregeo}

In a container of volume $\mathcal{V}$ that can be continuously deformed into a sphere, such that the height of the fluid column $h$ along the rotation axis is a homeomorphism between the volume $\mathcal{V}$ and the sphere,
all contours of constant $h$ (geostrophic contours) are closed. 
Examples of such containers include the full sphere (not a spherical shell) or ellipsoids. 
It is often postulated that incompressible flows in such a container can be expanded as \citep[e.g.][]{Greenspan1968}
\begin{equation}
    \vel = \sum_j^{\infty} \gamma_j(t) \, \vel_{G,j}(\posp) + \sum_i^\infty \alpha_i(t) \, \vel_i (\pos), \label{eq:inertialdecomp}
\end{equation}
where $\vel_{G,j}(\posp)$ are the (degenerate) geostrophic solutions \citep[e.g.][in spheres]{liao2010new} that only depend on the position perpendicular to the rotation axis $\posp$. 
They are given by the geostrophic equilibrium
\begin{equation}
    2\, \vec{\Omega} \times \vel_{G,j} = -\nabla p_{G,j},
    \label{eq:geobalance}
\end{equation}
and their superposition is commonly referred to as the geostrophic mode $\vel_G=\sum_j\vel_{G,j}$ \citep[e.g.][]{Greenspan1968}.
Additionally, $\vel_i (\pos)$ are the spatial eigensolutions of the inertial wave equation \citep[e.g.][ in ellipsoids]{Vantieghem2014}
\begin{equation}
    \frac{\partial\vel_i}{\partial t} + 2\, \vec{\Omega}\times\vel_i = -\nabla p_i.
    \label{eq:inertialwaveeq}
\end{equation}
Expansion (\ref{eq:inertialdecomp}) has proven to be exact for the ellipsoid \citep{backus2017completeness,ivers_enumeration_2017}.

From balance \eqref{eq:geobalance} it is clear that the axial geostrophic pressure torque vanishes, as the axial Coriolis torque vanishes for any flow $\vel$.
However, this is no longer the case when the flow is time dependent, even if it remains mainly geostrophic \citep[or 'pseudo-geostrophic',][]{gans-1971}, such that $\vel_{PG}(\posp,t) \simeq \sum_j\gamma_j(t) \, \vel_{G,j}$ (i.e. with $|\gamma_j| \gg |\alpha_i|$). In the presence of a Lorentz force the pseudo-geostrophic flow is governed by
\begin{equation}
    \frac{\partial\vel_{PG}}{\partial t} =-\frac{2}{\Le}\, \rotax\times\vel_{PG} -\nabla p + \left(\boldsymbol{\nabla}\times\B\right)\times\B. \label{eq:pseudogeo1}
\end{equation}
Using the geostrophic equilibrium \eqref{eq:geobalance} we substitute the Coriolis acceleration for its pressure gradient. Additionally, rewriting the Lorentz force in terms of the magnetic pressure gradient and the Maxwell term, \eqref{eq:pseudogeo1} takes the form
\begin{equation}
    \frac{\partial\vel_{PG}}{\partial t} = -\nabla (p_A+p_m) + (\B\boldsymbol{\cdot}\boldsymbol{\nabla})\B,
\end{equation}
with $p_A=p-\sum_j\gamma_jp_{G,j}$ and $p_m=\B^2/2$.
Besides the magnetic pressure $p_m$, an ageostrophic component $p_A$ remains in the pressure. They may both exert a torque on the container if it is not spherical.

\begin{figure}
  \begin{center}
  \includegraphics[width=\imwidth]{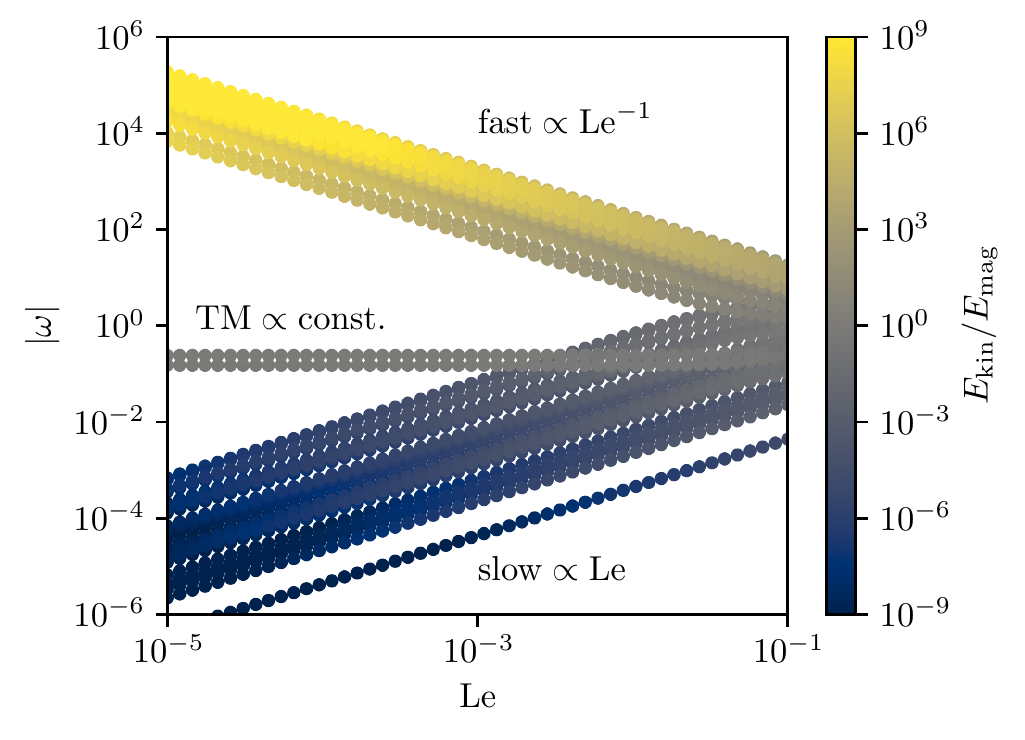}
  \caption{Mode frequencies as a function of Lehnert number in the sphere. The imposed magnetic field is $\B_0 =(-y,x-z/10,x/10)^\mathsf{T}$, following \citet{vidal2019binaries}. 
  The colours indicate the ratio of kinetic energy to magnetic energy, where yellow indicates a larger kinetic energy and blue a larger magnetic energy. The modes are separated into slow modes, fast modes and TM for $\Le\ll 1$. \label{fig:modeintroduction}}
\end{center}
\end{figure}

TM, also called 'torsional oscillations' \citep{braginsky_torsional_1970}, are examples of such pseudo-geostrophic flows.
They are solutions of the linearised equations \eqref{eq:goveqadimideal} for $\Le\ll 1$, and reduce to the ordinary geostrophic mode in the limit $\Le \rightarrow 0$. 
When scaled by the reciprocal of the Alfv\'en time scale $T_A$, the TM frequencies are constant (see Figure \ref{fig:modeintroduction}). 
Their Alfv\'en wave nature is also evident in the ratio of kinetic energy to magnetic energy, which is $\mathcal{O}(1)$ as indicated by the grey colour in Figure \ref{fig:modeintroduction}. We define TM to have a frequency independent of $\Le$ when $\Le\ll 1$ (if scaled by $T^{-1}_A$) and of approximately unit ratio between kinetic and magnetic energy. 
These two features clearly differentiate them from other modes present, namely the so-called fast modes and slow modes. 
The fast modes are slightly modified inertial modes, with frequencies on the order of the angular frequency, and their energy is mostly kinetic (see Figure \ref{fig:modeintroduction}, yellow dots). 
The slow modes (or Magneto-Coriolis modes) have a frequency much lower than the angular frequency and a small kinetic energy compared to the magnetic energy (see Figure \ref{fig:modeintroduction}, dark blue dots).

In the axisymmetric case, the geostrophic mode can be written as $\vel_G=u_G(s)\vec{1}_\phi$ and a pseudo-geostrophic flow is simply $\vel_{PG}\simeq u_{PG}(s,t)\vec{1}_\phi$ (with $s$ the cylindrical radius and $\phi$ the azimuthal angle).
The projection of the linearised momentum equation \eqref{eq:momeqnondimND} onto the geostrophic mode reduces to the one-dimensional equation
\begin{equation}
    \rho h\frac{\partial^2 u_{PG}(s,t)}{\partial t^2} = \frac{1}{s^2}\frac{\partial}{\partial s}\left(h s^3\frac{\partial}{\partial s}\left(\frac{u_{PG}(s,t)}{s}\right)\int B_s^2 dz\right),
\end{equation}
only depending on the radial distance $s$ to the rotation axis. 
\citet{roberts_theory_2012} referred to this equation as the canonical torsional wave equation. 
We refer the reader to \citet{roberts_electromagnetic_1972} and \citet{Jault2003} for details on the derivation. 
In the case of the ellipsoid, we shall consider TM within the framework of a QG model retaining ageostrophic components of the flow.

\subsection{Quasi-geostrophic equation with generic geostrophic contours}

We assume that the horizontal velocity components are independent of the coordinate $z$ along the rotation axis, $\vel_\perp = \vel_\perp(\pos_\perp,t)$. 
Together with the non-penetration boundary condition, $\vel\cdot\vec{n}=0$ on $\partial\mathcal{V}$, the mass continuity equation $\boldsymbol{\nabla}\boldsymbol{\cdot}\vel=0$ and the assumption of an equatorially symmetric volume $\mathcal{V}$ the QG velocity takes the form \citep[e.g.][]{Bardsley2018}
\begin{equation}
    \qgvel=\vel_\perp(\posp,t) + u_z\ez=\nabla\psi\times\nabla\left(\frac{z}{h}\right), \label{eq:qgvel}
\end{equation}
with $h=h(\posp)$ the height of the fluid column, $\vel_\perp(\posp,t) = \frac{1}{h}\nabla\psi\times\ez$,  $u_z=\frac{z}{h}\vel_\perp\cdot\nabla h$ and $\psi=\psi(\pos_\perp,t)$ a scalar stream function. 
By construction, $\psi$ is constant at the equator $\partial\mathcal{A}$ for the volume $\mathcal{V}$ considered here. Following the boundary condition arising naturally when $h\rightarrow 0$ at $\partial\mathcal{A}$ \citep{maffei_characterization_2017}, we choose $\psi=0$ on $\partial\mathcal{A}$.
Note that, if $\psi$ is constant along geostrophic contours (i.e. it is a function of $h$ only), we recover the geostrophic velocity (see Appendix \ref{appendix:geostrophic}). 

To derive an evolution equation for this scalar stream function, we project the momentum equation \eqref{eq:momeqnondimND} onto the subset $\qgvel'$ of QG velocities \eqref{eq:qgvel} following \citet{Labbe2015} and \cite{Bardsley2018}. This method is essentially a variational approach, which consists in finding solutions $\qgvel$ satisfying
\begin{equation}
    \int_\mathcal{V} \qgvel'\cdot \vec{f}(\qgvel)\, \dV =0, \qquad \forall \qgvel'\label{eq:varapproach}
\end{equation}
where 
\begin{equation}
    \vec{f}(\qgvel)=\frac{\partial \qgvel}{\partial t}+(\qgvel\boldsymbol{\cdot}\boldsymbol{\nabla})\qgvel+\frac{2}{\Le}\rotax\times\qgvel+\nabla p -\left(\boldsymbol{\nabla}\times\B\right)\times\B,
\end{equation}
with $\qgvel'$ and $\qgvel$ of the form \eqref{eq:qgvel}. 
Substituting \eqref{eq:qgvel} into \eqref{eq:varapproach} yields
\begin{subequations}
\label{eq:inner_product}
\begin{align}
    \int_\mathcal{V} \qgvel'\cdot \vec{f}\, \dV &= \int_\mathcal{V} \nabla\psi'\times\nabla\left(\frac{z}{h}\right)\cdot \vec{f}\, \dV,\\
    &= \int_\mathcal{A} \nabla\psi' \cdot \left<\nabla\left(\frac{z}{h}\right)\times\vec{f}\right>\, \dA,\\
    &= -\int_\mathcal{A} \psi'\boldsymbol{\nabla}\boldsymbol{\cdot}\left<\nabla\left(\frac{z}{h}\right)\times\vec{f}\right>\, \dA,\\
    &= -\int_\mathcal{A} \psi'\Qop (\vec{f})\, \dA,\label{eq:projection}
\end{align}
\end{subequations}
with the projection operator $\Qop$ defined as

\begin{equation}
    \Qop(\vec{f})=\boldsymbol{\nabla}\boldsymbol{\cdot}\left<\nabla\left(\frac{z}{h}\right)\times\vec{f}\right>,
\end{equation}
where $\left<\cdot\right>=\int_{-h}^h\cdot\, \mathrm{d}z$ is the integral along the rotation axis and $\int_\mathcal{A}\cdot\, \dA$ the integral over the equatorial surface plane $\mathcal{A}$ (shown in Figure \ref{fig:aligned_ellipsoid} for the ellipsoid). 
In this step, we made use of the boundary condition $\psi=0$ at the equator $\partial\mathcal{A}$. 
For expression \eqref{eq:projection} to be zero for any test function $\psi'$, the QG velocity $\qgvel$ must satisfy
\begin{equation}
    \Qop\left(\frac{\partial\qgvel}{\partial t} + \qgvel\boldsymbol{\cdot}\boldsymbol{\nabla}\qgvel+\frac{2}{\Le}\rotax\times\qgvel -\left(\boldsymbol{\nabla}\times\B\right)\times\B\right) = 0,
\end{equation}
where the pressure gradient is omitted, as it vanishes in the projection.

First, we consider the inertial term, which simplifies as
\begin{subequations}
\begin{align}
    \Qop\left(\frac{\partial\qgvel}{\partial t}\right)&= \boldsymbol{\nabla}\boldsymbol{\cdot}\left<\nabla\left(\frac{z}{h}\right)\times \left(\nabla\frac{\partial \psi}{\partial t}\times\nabla\left(\frac{z}{h}\right)\right)\right>,\\
    &= 2\Dop\frac{\partial\psi}{\partial t},
\end{align}
\end{subequations}
with
\begin{align}
    \Dop\Psi &= \boldsymbol{\nabla}\boldsymbol{\cdot} \left(\frac{1}{h} \nabla\Psi + \frac{1}{3h}\nabla h\times \left( \nabla \Psi \times \nabla h \right) \right).\label{eq:dop}
\end{align}

We can derive the projection for a force in the form of $\boldsymbol{\xi}\times\qgvel$ as follows
\begin{align}
    \Qop\left(\boldsymbol{\xi}\times\qgvel\right)=\boldsymbol{\nabla}\boldsymbol{\cdot}\left<-\left(\nabla\left(\frac{z}{h}\right)\cdot\boldsymbol{\xi}\right)\qgvel\right>,
\end{align}
which holds for any $\qgvel$ satisfying the boundary condition $\qgvel\bcdot\vec{n}=0$ on $\partial\mathcal{V}$.
We may further simplify this by considering $\Phi =- \nabla\left(\frac{z}{h}\right)\cdot\boldsymbol{\xi}$
\begin{subequations}
\begin{align}
    \boldsymbol{\nabla}\boldsymbol{\cdot}\left<\Phi\qgvel\right> &= \boldsymbol{\nabla}\boldsymbol{\cdot} \left<\frac{\Phi}{h}\boldsymbol{\nabla}\times\ez + \Phi z \nabla\psi\times\nabla\left(\frac{1}{h}\right)\right>,\\
    &= \boldsymbol{\nabla}\boldsymbol{\cdot} \left(\frac{\left<\Phi\right>}{h} \nabla\psi\times\ez + \left<z\Phi\right>\nabla\psi\times\nabla\left(\frac{1}{h}\right)\right),\\
    &= \left\{\frac{\left<\Phi\right>}{h},\psi\right\},
\end{align}
\end{subequations}
with
\begin{equation}
    \left\{X,Y\right\} = \left(\nabla X\times \nabla Y\right) \cdot\ez.\label{eq:jacop}
\end{equation}

Let us write $\qgvel\cdot\nabla\qgvel = (\boldsymbol{\nabla}\times\qgvel)\times\qgvel+\nabla\qgvel^2/2$. Since the gradient term vanishes exactly in the projection, the non-linear term can be written in the generic form $\boldsymbol{\xi}\times\qgvel$, with $\boldsymbol{\xi}=\boldsymbol{\nabla}\times\qgvel$.
For the non-linear term we thus have
\begin{subequations}
\begin{align}
    \left<\Phi\right> &= \left<-\nabla\left(\frac{z}{h}\right)\cdot\left(\boldsymbol{\nabla}\times\qgvel\right)\right>,\\
    &= \left<\boldsymbol{\nabla}\boldsymbol{\cdot}\left( \nabla\left(\frac{z}{h}\right)\times\left(\nabla\psi\times\nabla\left(\frac{z}{h}\right)\right) \right)\right>,\\
    &= 2\Dop\psi.
\end{align}
\end{subequations}
We have used here 
$\left<\boldsymbol{\nabla}\boldsymbol{\cdot}\vec{x}\right>=\boldsymbol{\nabla}\boldsymbol{\cdot}\left<\vec{x}\right>$,
which can be demonstrated to hold for $\vec{x}=\nabla\left(\frac{z}{h}\right)\times\left(\nabla\psi\times\nabla\left(\frac{z}{h}\right)\right)$. 
The non-linear term is then given by
\begin{equation}
    \Qop\left((\boldsymbol{\nabla}\times\qgvel)\times\qgvel\right)=2\left\{\frac{1}{h}\Dop\psi,\psi\right\}.\label{eq:nonlinearterm}
\end{equation}

For the Coriolis force $\boldsymbol{\xi}=2/\Le\, \rotax$ and thus  $\left<\Phi\right> = \left<-\nabla\left(\frac{z}{h}\right)\cdot\vec{\Omega}\right> = -4/\Le$, so that the Coriolis force reduces to
\begin{equation}
   \Qop\left(\frac{2}{\Le}\rotax\times\qgvel\right)=-\frac{4}{\Le} \left\{\frac{1}{h},\psi\right\}.
\end{equation}

The QG scalar momentum equation is then given by

\begin{equation}
    \Dop\frac{\partial\psi}{\partial t} + \left\{\frac{1}{h}\Dop\psi,\psi\right\} =\frac{2}{\Le}\left\{\frac{1}{h},\psi\right\}+\frac{1}{2}\Qop\left(\left(\boldsymbol{\nabla}\times\B\right)\times\B\right).\label{eq:momeqhybrid}
\end{equation}
We can close the system by assuming that the 3-D magnetic field in \eqref{eq:momeqhybrid} is advected only by the QG velocity.
This is, to the authors' knowledge, the first presentation of a hybrid model with QG velocities and 3-D magnetic field. Such a model is desirable especially in geodynamo modelling, where it is found that a strong columnar motion is accompanied by a magnetic field of 3-D structure \citep[e.g.][]{schaeffer_turbulent_2017}.

To derive a fully 2D model we assume the same form for the magnetic field, as for the velocity
\begin{equation}
    \B=\nabla A\times\nabla\left(\frac{z}{h}\right), \label{eq:aformulation}
\end{equation}
with $A=A(\posp,t)$ a scalar potential. 
By construction, such a magnetic field satisfies the perfectly conducting boundary condition  $\B\boldsymbol{\cdot}\vec{n}=0$. 
This approximation has been used previously to investigate TM in QG models \citep{Canet2014,Labbe2015}. 
Under this assumption, the Lorentz term simplifies analogous to \eqref{eq:nonlinearterm}, such that
\begin{equation}
  \Qop\left(\left(\boldsymbol{\nabla}\times\B\right)\times\B\right)= 2\left\{\frac{1}{h}\Dop A, A\right\}.
\end{equation}
The scalar momentum equation is then written in terms of $\psi$ and $A$ only
\begin{equation}
    \Dop\frac{\partial\psi}{\partial t} + \left\{\frac{1}{h}\Dop\psi,\psi\right\} =2\Omega\left\{\frac{1}{h},\psi\right\}+ \left\{\frac{1}{h}\Dop A,A\right\}.\label{eq:momeqproject}
\end{equation}
The ideal induction equation \eqref{eq:indeqideal} can be simplified as follows
\begin{align}
    \qgvel\times\B &= \frac{1}{h^2}\nabla\psi\times\nabla A - \frac{z}{h^3}\left\{\psi,A\right\}\nabla h,\\
    &= \frac{C}{h}\ez - \frac{Cz}{h^2}\nabla h,
\end{align}
with $C = \frac{1}{h}\left\{\psi,A\right\}$. Taking the curl then gives
\begin{subequations}
\begin{align}
    \boldsymbol{\nabla}\times\left(\qgvel\times\B\right) &= \frac{1}{h}\nabla C\times\ez - \frac{z}{h^2} \nabla C\times\nabla h\\
    &=\nabla C\times\nabla\left(\frac{z}{h}\right).
\end{align}
\end{subequations}
Thus, the induction equation is given by
\begin{align}
    \frac{\partial A}{\partial t} = \frac{1}{h}\left\{\psi,A\right\}.
    \label{eq:indeqproject}
\end{align}
In the sphere, where cylindrical coordinates apply, the equations \eqref{eq:momeqproject} and \eqref{eq:indeqproject} are exactly equivalent to the equations obtained by \citet{Labbe2015}.

\section{Methods for the ellipsoid}

\begin{figure}
  \begin{center}
   \begin{overpic}[width=\imwidth,unit=1mm]{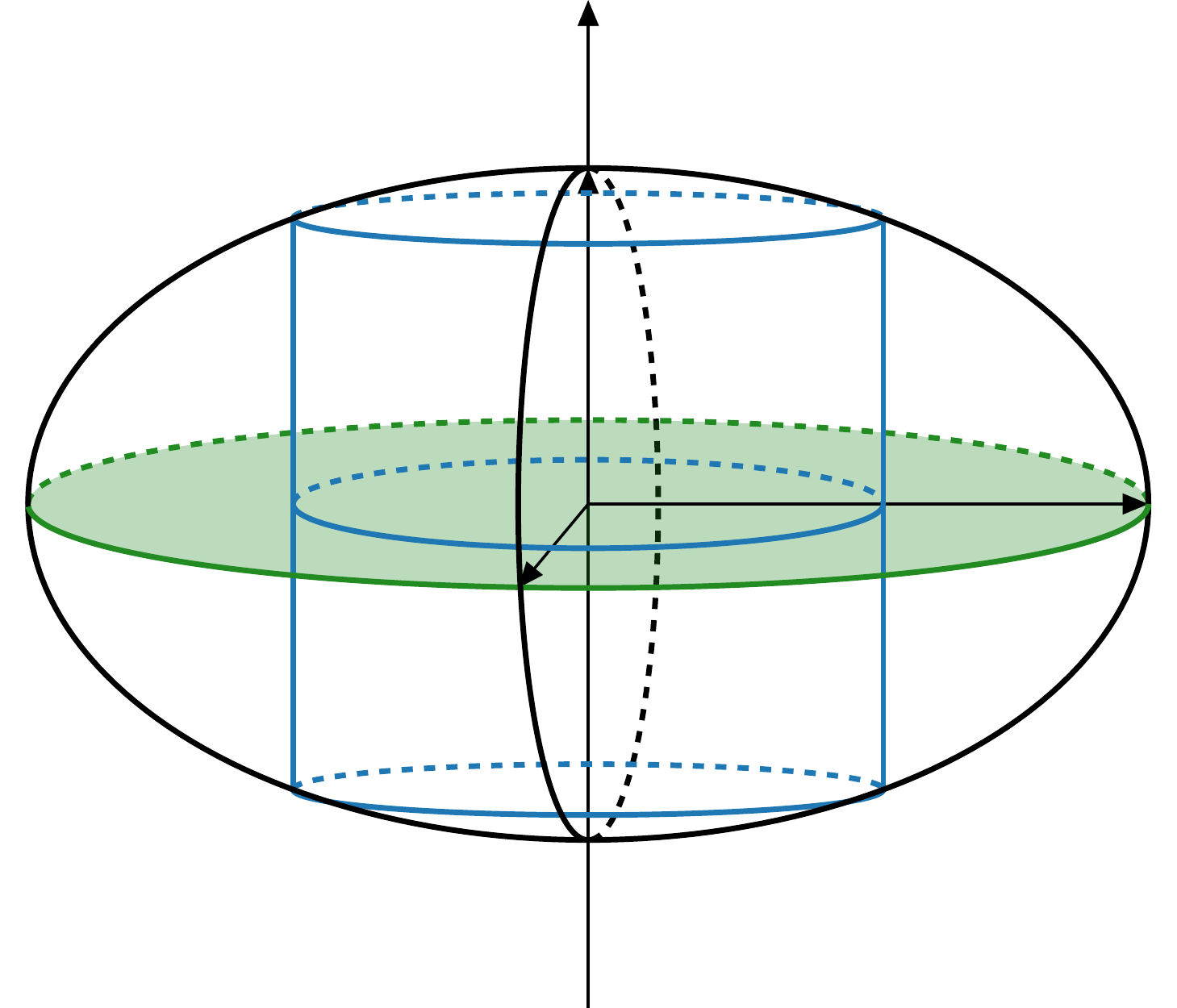}
    \put(52,80){\Large $\vec{\Omega}$}
    \put(80,55){\Large $\mathcal{V}$}
    \put(85,63){\Large $\partial \mathcal{V}$}
    \put(15,42){\Large \color{forestgreen} $\mathcal{A}$}
    \put(15,32){\Large \color{forestgreen} $\partial\mathcal{A}$}
    \put(51,55){\large $c$}
    \put(47,36){\large $b$}
    \put(80,40){\large $a$}
\end{overpic}
  \caption{Schematic of a geostrophic column (blue) in an ellipsoid of volume $\mathcal{V}$ where one of the principal axes is aligned with the rotation axis $\vec{\Omega}$. The area of the equatorial plane $\mathcal{A}$ is shaded in green.}
  \label{fig:aligned_ellipsoid}
  \end{center}
\end{figure}
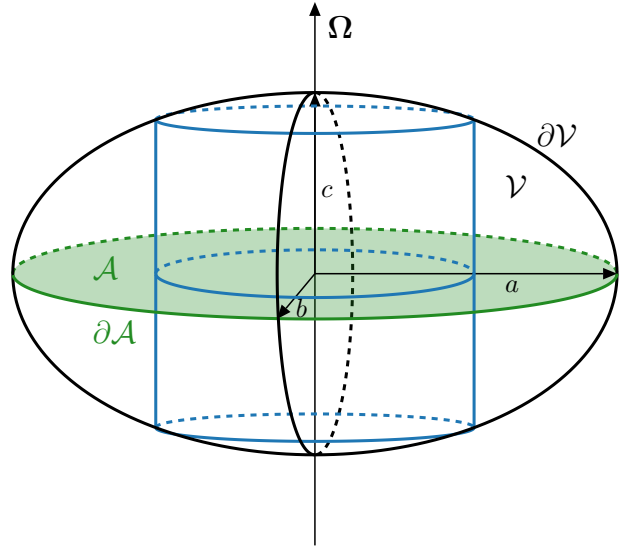

We now consider the case of an ellipsoid with semi axes $a$, $b$ and $c$ defined by
\begin{equation}
    \frac{x^2}{a^2}+\frac{y^2}{b^2}+\frac{z^2}{c^2} = 1.
\end{equation}
To keep equatorial symmetry, we also consider that the rotation axis is aligned with the $c$-axis, $\rotax=\ez$ (Figure \ref{fig:aligned_ellipsoid}).

\subsection{Cartesian monomial basis in the ellipsoid}
\label{appendix:cartesian}
Since the ellipsoid is a quadratic surface,  smooth-enough solutions can be sought by using an infinite sequence of Cartesian polynomials \citep{Lebovitz1989}. 
This approach has proven accurate to describe 3-D inviscid flows in ellipsoids \citep[e.g.][]{vantieghem2015latitudinal,Vidal2017,vidal2020compressible}. 
The 3-D inertial modes are exactly described by polynomials in the ellipsoid \citep{backus2017completeness,ivers_enumeration_2017}, and also the QG and 3-D inertial modes in the spheroid \citep{maffei_characterization_2017,zhang2017theory}. 
Additionally, the MHD modes upon idealised background magnetic fields \citep[e.g.][]{malkus_hydromagnetic_1967} also have an exact polynomial description in the spheroid \citep{kerswell1994tidal} and the ellipsoid \citep{vidal2017diffusionless}. 

Similarly, a 2-D polynomial decomposition in the Cartesian coordinates can be obtained for arbitrary QG vector (\ref{eq:qgvel}) in non-axisymmetric ellipsoids as follows. 
To satisfy the polynomial form of the velocity components, the stream function must be given as
\begin{equation}
  \psi = h^3 \, \Pi(x,y) = h^3 \sum_{i}\hat{\alpha}_{i}\Pi_{i},
\end{equation}
with the complex-valued coefficients $\hat{\alpha}_{i}$ and the monomials 
\begin{equation}
    \Pi_i = 1,x,y,xy,x^2, ...,x^{N-1},y^{N-1}
\end{equation}
with $i \in [0, N_2]$ and $N_2=N(N+1)/2$. 
At any point $(x,y)$ we have
\begin{equation}
  \frac{h^2}{c^2} = 1-\frac{x^2}{a^2}-\frac{y^2}{b^2}.
\end{equation}
If additionally we define $G=h^2/2$, we can rewrite
\begin{equation}
  h\nabla h = \nabla G.\label{eq:rewritegradh}
\end{equation}
Then, the QG basis vectors $\qgvel_i$ are given by
\begin{equation}
  \qgvel_i = h^2\nabla\Pi_i\times\ez + 3\Pi_i\nabla G\times\ez-z\nabla\Pi_i\times\nabla G,
  \label{eq:qg-basis}
\end{equation}
with the first three basis elements
\begin{subequations}
\label{eq:firstbasisvecs}
\begin{align}
\qgvel_0&=3c^2\begin{pmatrix}-y/b^2\\x/a^2\\0\end{pmatrix},\\ 
\qgvel_1&=c^2\begin{pmatrix}1-x^2/a^2-4y^2/b^2\\3xy/a^2\\-xz/a^2\end{pmatrix},\\
\qgvel_2&=c^2\begin{pmatrix}-3xy/b^2\\4x^2/a^2+y^2/b^2-1\\yz/b^2\end{pmatrix}.
\end{align}
\end{subequations}
The full velocity is reconstructed by
\begin{equation}
  \qgvel = \sum_{i=0}^{N_2}\hat{\alpha}_i\qgvel_i.\label{eq:eigenvel}
\end{equation}

For the linear hydrodynamic (Rossby wave) problem
\begin{equation}
    \Dop\frac{\partial\psi}{\partial t} = 2\Omega\, \left\{\frac{1}{h},\psi\right\},
\end{equation}
the polynomial degree of $\psi_i=\hat\alpha_i h^3\Pi_i$ is preserved, that is the QG inertia and Coriolis operators do not modify (increase) the polynomial degree, similar to the 3-D Coriolis operator in the ellipsoid \citep{backus2017completeness,ivers_enumeration_2017}. 
This is no longer the case in the presence of a background magnetic field within the QG model 
\citep[unless the magnetic field is only linear in the spatial coordinates, see][]{malkus_hydromagnetic_1967},
as the Lorentz term modifies the polynomial degree. Then, the exact solutions cannot be obtained from a finite set of $\Pi_i$. Hence, we must project the governing equations onto the basis with a sufficiently large maximum polynomial degree. 

\subsection{Galerkin method}\label{sec:galerkin}

Since we are interested in the wave properties, we linearise equations (\ref{eq:goveqadimideal}) around a background state with no motion and steady magnetic field $\B_0$. 
In the Earth's core, the characteristic mean velocity field is thought to be negligible compared to the Alfv\'en wave velocity \citep{gillet_planetary_2015,barenzung_modeling_2018}. 
Hence, the velocity and magnetic perturbations $[\tilde\vel,\tilde\B]$ are given by 
\begin{subequations}
\label{eq:goveqadimLN}
\begin{align}
    \begin{split}
    \frac{\partial \tilde\vel}{\partial t} + \frac{2}{\Le}\, \rotax\times\tilde\vel =& -\nabla p + (\boldsymbol{\nabla}\times\B_0)\times\tilde\B \\ 
    &+ (\boldsymbol{\nabla}\times\tilde\B)\times\B_0, \end{split} \label{eq:linmom}\\
    \frac{\partial \tilde\B}{\partial t} =& \boldsymbol{\nabla}\times\left(\tilde\vel\times\B_0\right).\label{eq:linind}
\end{align}
\end{subequations}
The linearised set of equations in the hybrid model then read
\begin{subequations}
\label{eq:linhybrideq}
\begin{align}
    \begin{split}
    \Dop\frac{\partial\tilde\psi}{\partial t} =&\frac{2}{\Le}\left\{\frac{1}{h},\tilde\psi\right\}+\frac{1}{2}\Qop((\nabla\times\B_0)\times\tilde\B\\
    &+\frac{1}{2}\Qop((\nabla\times\tilde\B)\times\B_0),\end{split}\\
    \frac{\partial\tilde\B}{\partial t} =& \nabla\times\left(\left(\nabla\tilde\psi\times\nabla\left(\frac{z}{h}\right)\right)\times\B_0\right).
\end{align}
 \end{subequations}
The linearisation of the magnetic field translates to $A=A_0+\tilde{A}$ for the scalar potential and the scalar QG equations \eqref{eq:momeqproject} and \eqref{eq:indeqproject} read
\begin{subequations}
\label{eq:lineqqg}
\begin{align}
    \partial_t \Dop\tilde\psi-\frac{2}{\Le}\left\{\frac{1}{h},\tilde\psi\right\} &=\left\{\frac{1}{h}\Dop \tilde{A},A_0\right\} + \left\{\frac{1}{h}\Dop A_0,\tilde{A}\right\},\label{eq:momeqscalarlin}\\
    \frac{\partial \tilde A}{\partial t} &= \frac{1}{h}\lbrace \tilde\psi, A_0\rbrace.\label{eq:indeqscalarlin}
\end{align}
\end{subequations}

To solve such sets of linearised equations for eigenmodes, Fourier expansions along the azimuthal direction could be used in the sphere, combined with finite differences in the radial direction \citep[][]{Labbe2015}.
Here, we use a Galerkin approach to project the governing equations onto the respective polynomial bases \citep[e.g.][]{Vidal2017,vidal2020compressible}. 
This approach is suitable for the Cartesian monomial basis, as we can analytically integrate the Cartesian monomials occurring in the inner product \citep[see formula 50 in][]{Lebovitz1989}.
For the QG model this projection is given by
\begin{equation}
    f_{ij}= \int_\mathcal{A} \tilde\psi_i f(\tilde\psi_j,\tilde A_j)\, \dA, \label{eq:qgprojection}
\end{equation}
where now $f(\tilde\psi,\tilde A)$ corresponds to a force in the scalar momentum equation \eqref{eq:momeqscalarlin}. In this way we create coefficient matrices $U_{ij}$, $C_{ij}$ and $L_{ij}$ for the inertial, Coriolis and Lorentz force, respectively. 
 Analogously, the induction equation \eqref{eq:indeqscalarlin} is projected onto the basis $A_i=\hat\zeta_ih^3\Pi_i$ and the coefficient matrices $B_{ij}$ and $V_{ij}$ correspond to the projections of the temporal change of the magnetic field and magnetic advection, respectively. 
 For this model $U_{ij}$ and $B_{ij}$ are identical and Hermitian. Assuming that $\tilde\psi(\posp,t) = \hat\psi(\posp)\exp(\mathrm{i}\omega t)$ (and the same for $\tilde A$), so that $\partial_t\tilde\psi=\mathrm{i}\omega\tilde\psi$, the resulting matrix form is
\begin{equation}
    \mathrm{i} \omega\vec{M}\vec{x}=\vec{D}\vec{x},
\end{equation}
with $\vec{M},\vec{D}\in \R^{2N_2\times2N_2}$ of the form
\begin{align}
    \vec{M}=\begin{pmatrix}
    U_{ij} & 0\\
    0 & B_{ij}
    \end{pmatrix}, && \vec{D}=\begin{pmatrix}
    C_{ij} & L_{ij}\\
    V_{ij} & 0
    \end{pmatrix},
\end{align}
and $\vec{x}=(\hat\alpha_j,\hat\zeta_j)\in \Cn^{2N_2}$. 
This form is referred to as a generalised eigen problem solvable for eigen pairs $(\omega_k,\vec{x}_k)$.

Note that using the reduced equation and projecting onto the basis of stream functions $\tilde\psi_i$ is equivalent to projecting the 3-D equations onto the QG basis $\qgvel_i$, apparent from \eqref{eq:inner_product}. 
We use this fact for the hybrid model and project the 3-D momentum equation \eqref{eq:linmom} onto the QG basis vectors $\qgvel_i$ while keeping the full 3-D basis vectors $\B_i$ with coefficients $\zeta_i$ for the magnetic field. The induction equation \eqref{eq:linind} is projected onto the basis $\B_i$.
The resulting matrices are $U'_{ij},C'_{ij}\in\R^{N_2\times N_2}$, $L'_{ij}\in\R^{N_2\times N_3}$, $B_{ij}\in\R^{N_3\times N_3}$ and $V'_{ij}\in\R^{N_3\times N_2}$, so that $\vec{M}',\vec{D}'\in \R^{N_2+N_3\times N_2+N_3}$ and $\vec{x}=(\hat{\alpha}_j,\zeta_j)\in \Cn^{N_2+N_3}$. 
These matrices can be built analytically, but this becomes tedious even for a maximum polynomial degree as low as 2 and in practise this is done by computer algebra systems or numerically.

\subsection{Numerical implementation}
\label{sec:numericaltools}

The linear problems based on Cartesian monomials are implemented in the Julia programming language \citep{bezanson_julia_2017}. 
The QG, hybrid and 3-D models are freely available at \url{https://github.com/fgerick/Mire.jl}. The reproduction of all the results and figures from this article using these models is available through \url{https://dx.doi.org/10.5281/zenodo.3631244}.

To solve for the eigen problems, different methods have been employed. To calculate the full spectrum of eigensolutions, we use either LAPACK or recent Julia implementations for accuracy beyond standard floating point numbers (e.g. in Figure \ref{fig:torques_z} below). 
Full spectrum eigensolutions are computationally demanding, which is why we also apply targeted iterative solvers from the ARPACK library, making use of the sparsity of the matrices $\vec{M}$ and $\vec{D}$, where approximately 13\% and 30\% of entries are non-zero, respectively.
The sparse solver is also applied to follow eigenbranches (i.e to track a specific eigensolution through the parameter space). To do so, we apply a targeted shift-and-invert method \citep[e.g.][]{rieutord1997inertial,vidal2015quasi}
\begin{equation}
    (\vec{D}-\sigma\vec{M})^{-1}\vec{D}\vec{x}=\lambda\vec{x}
\end{equation}
around a target $\sigma\in\Cn$, with the new eigenvalue $\lambda = (\mathrm{i} \omega-\sigma)^{-1}$. 
This strategy is efficient to compute the eigenvalues close
to the target $\sigma$ (which is chosen close to the desired eigenvalue $\mathrm{i} \omega$).

\section{Numerical results}

We first validate our QG (and hybrid) model against the 3-D model for a simplified background magnetic field \citep{malkus_hydromagnetic_1967}, and then consider a more complex background magnetic field that is able to drive TM. In this section the QG model is considered and we compare our results to a 3-D magnetic field with the hybrid model in Appendix \ref{appendix:hybrid}.

\subsection{Modes in the Malkus field}

\begin{figure}
  \begin{center}
  \includegraphics[width=\imwidth]{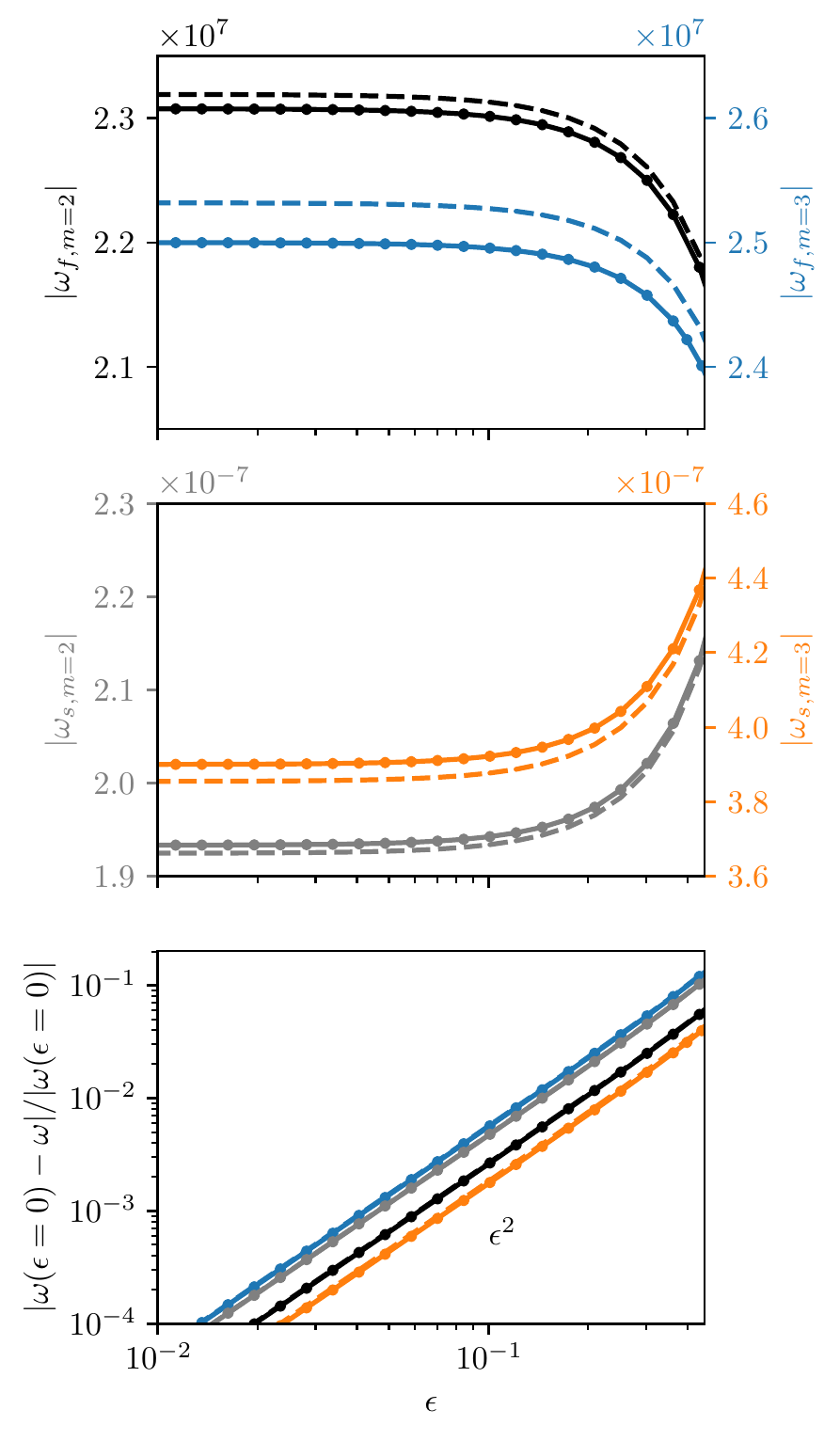}
  \caption{Frequencies $\omega_{f,m}$ of the fast modes (top) and $\omega_{s,m}$ of the slow modes (middle) as a function of ellipticity for radial complexity $l=1$ and azimuthal wave number $m=2,3$. Difference between the frequency as a function of $\epsilon$ and the frequency in the sphere with $\epsilon=0$, normalised by the frequency in the sphere, for the considered fast and slow modes (bottom). The Lehnert number is $10^{-8}$. The different models are: QG (solid), hybrid (dots) and 3-D (dashed). \label{fig:slowfast}}
\end{center}
\end{figure}

An interesting first study case is the mean field introduced by \citet{malkus_hydromagnetic_1967}, originally given as a field of uniform current along the rotation axis in a sphere with $\B_{0,M} = s \, \vec{1}_\phi$ (hereafter Malkus field). In his study, 
the slow and the fast modes were recovered from the resulting dispersion relation \citep[see eq. 2.28 in][]{malkus_hydromagnetic_1967}. 

In the ellipsoidal case, the Malkus field is modified to follow the elliptical geostrophic contours.
This translates into the background magnetic field $\vec{B}_{0,z} = c^2(-y/b^2, x/a^2,0)^\mathsf{T}$ in Cartesian coordinates \citep[e.g.][]{vidal2019binaries} and a mean magnetic potential $A_0=h^3/3$ for the QG model. Due to the lack of any magnetic field component perpendicular to the geostrophic contours, the Malkus field does not permit TM. 
However, that field allows us to investigate the slow and fast modes in the ellipsoid. We report, for the first time, the dependency of these modes on the equatorial ellipticity
\begin{equation}
  \epsilon = \frac{a^2-b^2}{a^2+b^2},
\end{equation}
where $\epsilon = 0$ corresponds to the axisymmetric case. Here, we investigate the parameter range $\epsilon \in [0,0.4]$. 
For all the results shown below, the semi-axis along the rotation axis is kept constant at $c=1$. 
The influence of polar flattening has already been investigated previously and is not discussed here \citep{maffei_characterization_2017,zhang2017theory}. 
Throughout this study, the volume is preserved by setting $a=1/b$ when increasing the ellipticity in the equatorial plane, so that $abc=1$.

We compute the frequencies of two of the largest-scale fast and slow modes, and track their frequencies as a function of $\epsilon$. 
The results are shown in Figure \ref{fig:slowfast}. The trends of all models agree well. 
The fast modes decrease in frequency, whereas the slow modes increase their frequency as the ellipticity is increased. 
The frequency is almost independent of the ellipticity when $\epsilon\ll 1$ , and the difference with respect to the spherical values scales as $|\omega(\epsilon=0)-\omega(\epsilon)|\sim \epsilon^2$ for the fast and slow modes (see Figure \ref{fig:slowfast}, bottom). This scaling may be anticipated by the relation of the fast and slow modes to the inertial modes in the ellipsoid, showing a similar scaling \citep[compare with equation 3.24 in ][]{Vantieghem2014}.

The Malkus field is completely determined by the geostrophic basis (as introduced in Appendix \ref{appendix:geostrophic}). 
Hence, we do not observe any differences between the QG model (solid line) and the hybrid model (dots). The differences in frequency magnitude between the 3-D model (dashed line) and the QG and hybrid model depend on the modes' complexity \citep[see][]{Labbe2015,maffei_characterization_2017}.
The discrepancies observed between the different models are similar over the entire range of ellipticities considered here ($0\leq \epsilon \lesssim 0.4$). We are thus confident in using the QG (or hybrid) models for further analysis, as we do not observe strong 3-D effects on the modes by the equatorial ellipticity. 

\subsection{Torsional Alfv\'en modes}

\begin{figure}
    \centering
    \includegraphics[width=\imwidth]{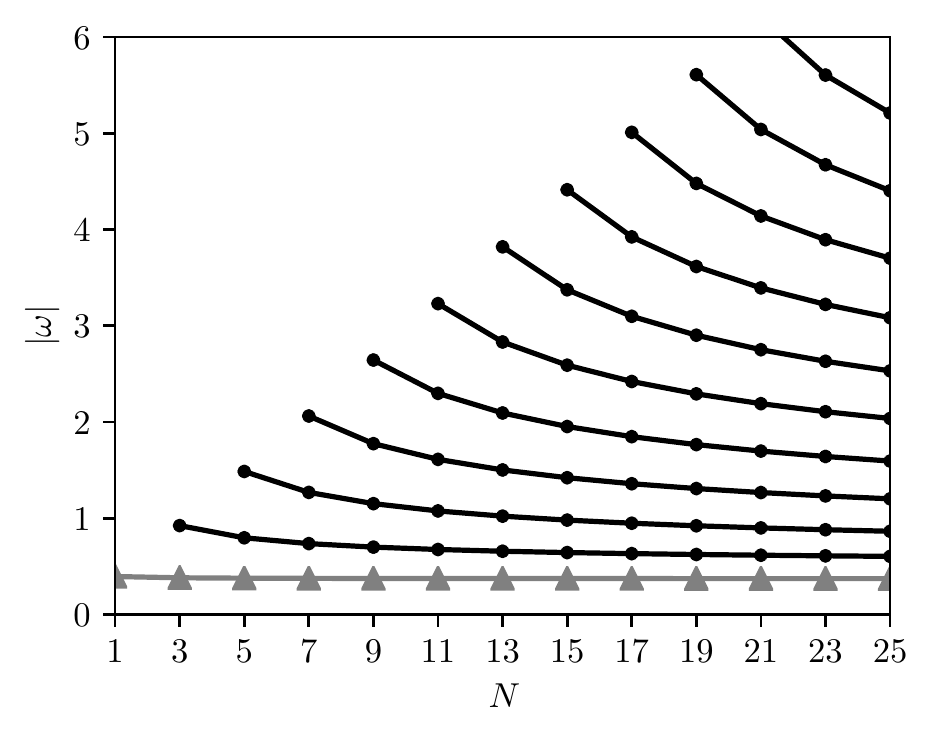}
    \caption{Convergence of frequencies as a function of truncation degree $N$ for a background magnetic field $\B_{0,\mathrm{QG}}$, $\epsilon=0.42$ and $\Le=10^{-5}$. Connected lines indicate individual modes. Black lines correspond to TM and the smallest frequency mode (in grey with triangles) is the $U_3$-mode.}
    \label{fig:convergence_torsionalfreq}
\end{figure}

To drive TM the imposed background magnetic field must have a component perpendicular to the geostrophic contours.
For the QG model, we must consider a scalar potential $A_0$ that is not only a function of $h$. We choose $A_0=h^3(1+x)/3$, which yields

\begin{equation}
    \B_{0,\mathrm{QG}} = \frac{c^2}{3}\begin{pmatrix}-3(1+x)y/b^2\\(3+4x)x/a^2+y^2/b^2-1\\yz/b^2\end{pmatrix}. \label{eq:aformb}
\end{equation}
Since the components of such a magnetic field are no longer linear in the Cartesian coordinates (contrary to the Malkus field), the convergence of the modes depends on the truncation of the maximum polynomial degree. 
We verify the convergence of the largest scale TM (see black lines in Figure \ref{fig:convergence_torsionalfreq}). As $N$ is increased more TM with a larger polynomial complexity appear, with one additional TM per two polynomial degrees. 
This is explained by the introduction of an additional geostrophic basis vector at every second polynomial degree (see \cite{backus2017completeness}, in the sphere and Appendix \ref{appendix:geostrophic} in the ellipsoid). 

\begin{figure}
    \centering
    \includegraphics[width=0.49\imwidth]{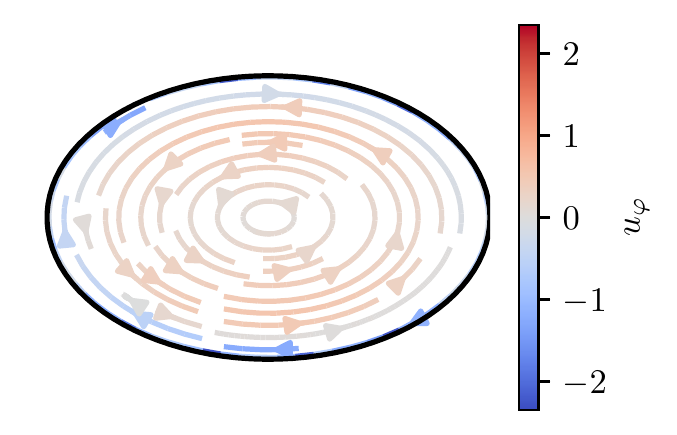}
    \includegraphics[width=0.49\imwidth]{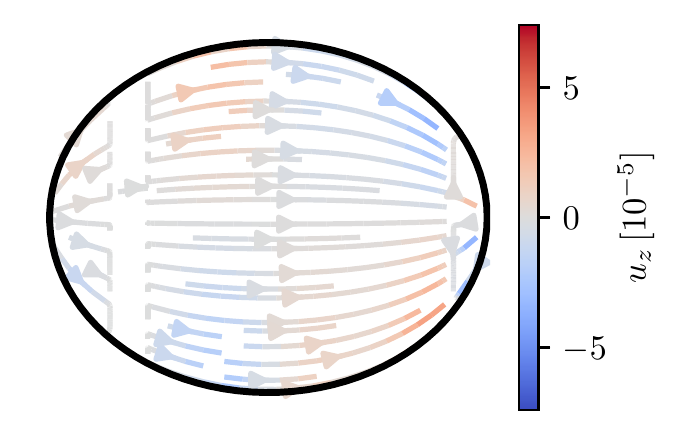}    \includegraphics[width=0.49\imwidth]{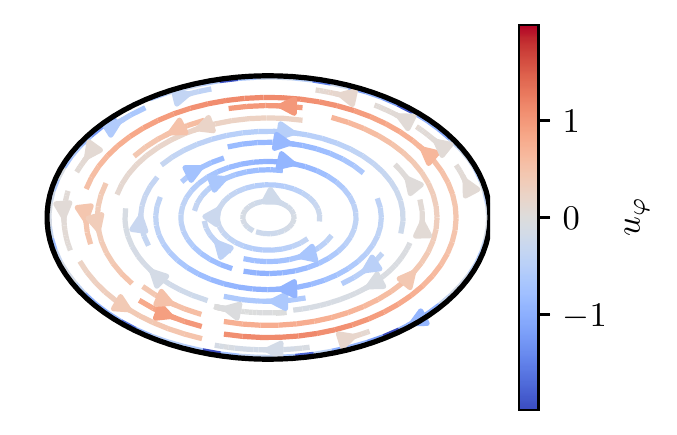}
    \includegraphics[width=0.49\imwidth]{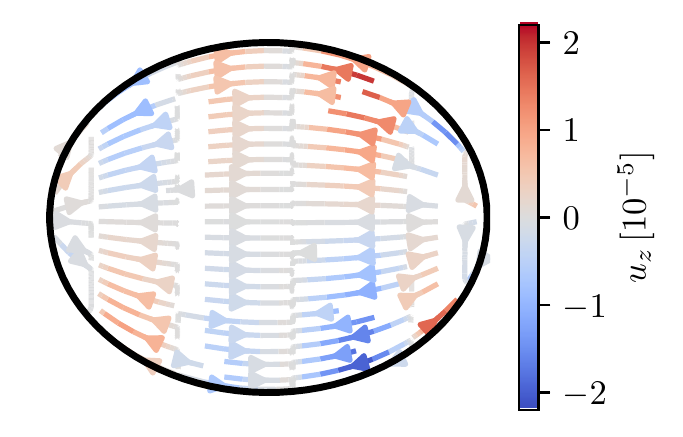}  
    \caption{Equatorial sections (left) and meridional sections along the $x$-axis (right) of the two largest scale TM using $\B_{0,\mathrm{QG}}$, $\epsilon=0.42$ and $\Le=10^{-5}$. The colours indicate the velocity along the geostrophic contours $u_\varphi$ and the vertical velocity $u_z$, respectively.}
    \label{fig:streamplots}
\end{figure}

The equatorial and meridional sections of the two lowest frequency (and thus largest scale) TM, calculated using $\B_{0,\mathrm{QG}}$ at $N=7$, are presented in Figure \ref{fig:streamplots} for a strongly deformed ellipsoid with equatorial ellipticity $\epsilon=0.42$. 
As in the sphere, the velocities of TM follow the geostrophic contours that are now ellipses. 
The velocity structure is almost purely horizontal, seen by the ratio of the velocity amplitudes $u_\varphi/u_z\sim 10^5$, where $u_\varphi$ is the velocity along an elliptical geostrophic contour and $u_z$ is the vertical velocity.

\begin{figure}
    \centering
    \includegraphics[width=0.49\imwidth]{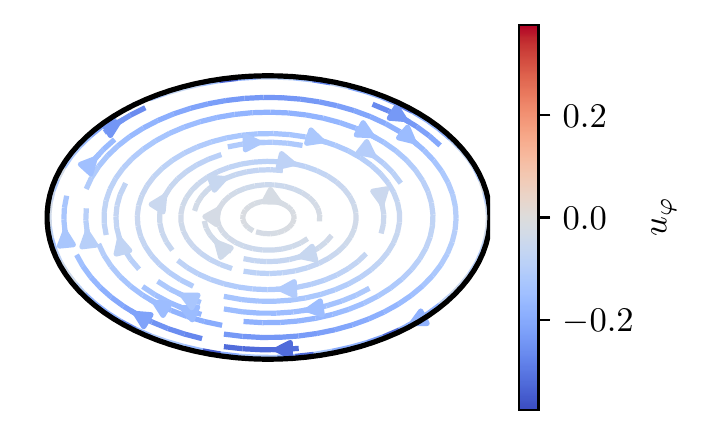}
    \includegraphics[width=0.49\imwidth]{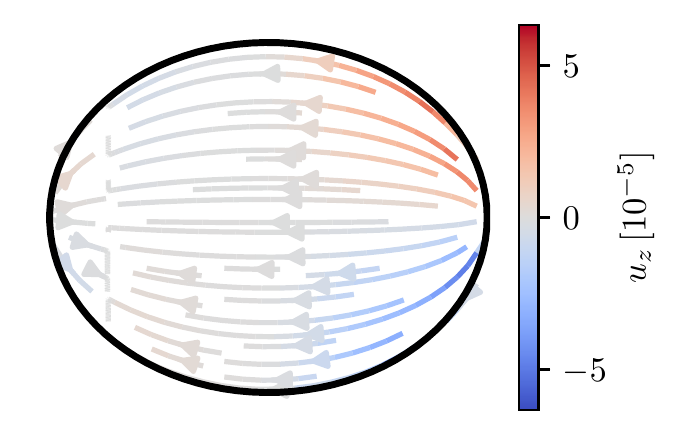}
    \caption{Equatorial section (left) and meridional section along the $x$-axis (right) of the $U_3$-mode using $\B_{0,\mathrm{QG}}$, $\epsilon=0.42$ and $\Le=10^{-5}$. The colours indicate the velocity along the geostrophic contours $u_\varphi$ and the vertical velocity $u_z$, respectively.}
    \label{fig:streamplot-u3}
\end{figure}

The lowest frequency mode (highlighted in grey triangles in Figure \ref{fig:convergence_torsionalfreq}) is hereafter referred to as $U_3$-mode. It is already present for a truncation degree $N=1$, where only components linear in the Cartesian coordinates are included. 
The equatorial and meridional section of the $U_3$-mode are presented in Figure \ref{fig:streamplot-u3} for an ellipsoid with equatorial ellipticity of $\epsilon=0.42$ and $N=7$. Compared to other TM, it consists almost solely of a velocity with uniform vorticity along the $z$ direction.

\subsubsection{Identification of torsional Alfv\'en modes}\label{sec:identification}

\begin{figure}
  \begin{center}
  \includegraphics[width=\imwidth]{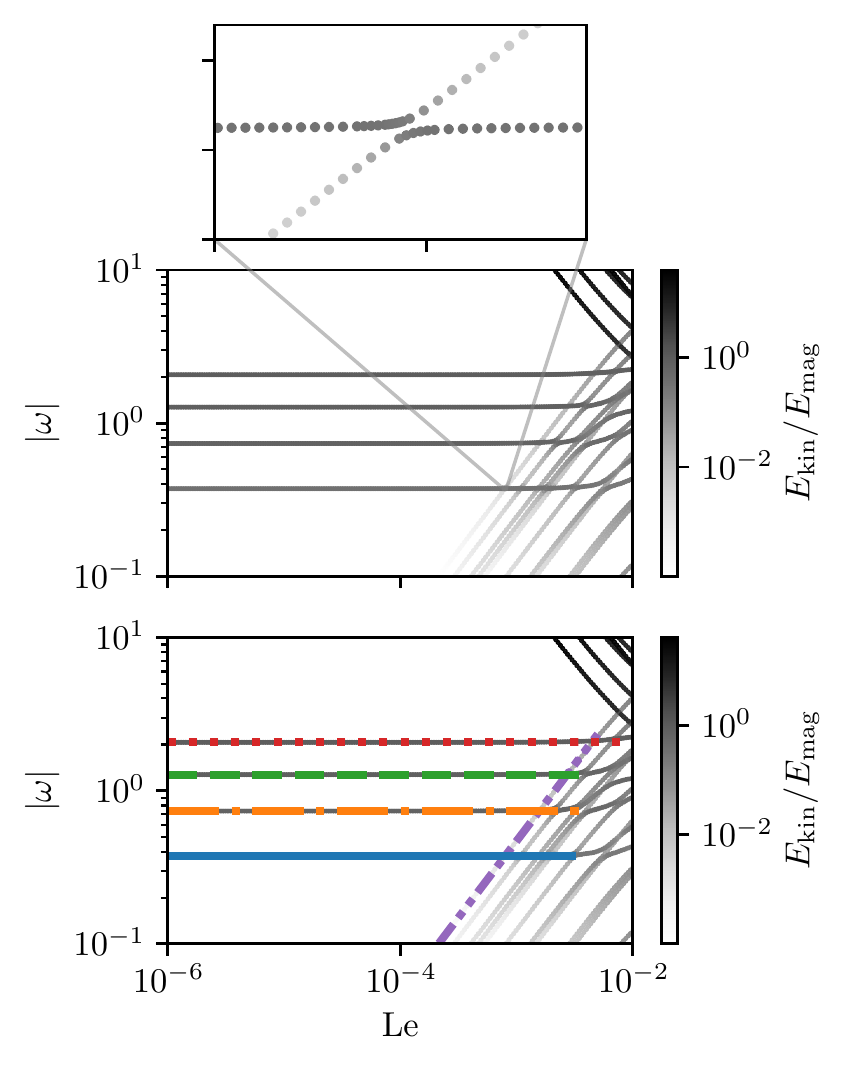}
  \caption{Frequency of eigensolutions as a function of Lehnert number for $\epsilon=0.42$, with colours indicating the ratio of kinetic to magnetic energy (top). 
  The inset shows an avoided crossing of the $U_3$-mode and the fastest slow mode, with dots indicating individual steps of the tracking algorithm. The frequencies of TM (orange dash-dotted, green dashed and red dotted), the $U_3$-mode (blue solid) and the fastest slow mode (purple dash-dot-dotted) are highlighted in the bottom Figure. \label{fig:avoidedcrossing}}
\end{center}
\end{figure}

When the Lehnert number is not sufficiently small to separate the branches of eigensolutions, as seen for the sphere in Figure \ref{fig:modeintroduction} at $\Le>10^{-3}$, a clear identification of TM in the spectrum of eigensolutions is complicated.
In Figure \ref{fig:avoidedcrossing} we show the dependency of the frequency of the eigensolutions on the Lehnert number for an ellipsoid with $\epsilon=0.42$.
For the TM represented in this Figure, no dependency of the frequency on $\Le$ is observed for $\Le\lesssim 10^{-3}$, as in the case of the sphere (compare Figure \ref{fig:modeintroduction}, bottom). Similarly, the $U_3$-mode shows no dependency of its frequency on $\Le$ for $\Le < 7\times 10^{-4}$. 
For $\Le\lesssim 10^{-3}$ the TM shown here and the $U_3$-mode do not cross any other eigensolutions (and due to their independence of $\Le$ they do not cross each other). In this region we have no difficulty in identifying individual TM or the $U_3$-mode.
When $\Le$ is increased to values greater than $10^{-3}$, more eigensolutions with frequencies close to the TM or the $U_3$-mode exist. Tracking these eigensolutions as a function of $\Le$ (as described in Section \ref{sec:numericaltools}) reveals that they can undergo so-called avoided crossings, where two eigensolutions approach each other without ever degenerating.
An example of such an avoided crossing is shown in the inset in Figure \ref{fig:avoidedcrossing}, where the $U_3$-mode morphs into the fastest slow mode and vice versa. 
The two modes exchange their properties, as shown here by the ratio of kinetic to magnetic energy. 
Such a behaviour has been similarly observed in other geophysical wave studies \citep{rogister_influence_2009}, even for non-vanishing diffusivities \citep{triana_coupling_2019}, or in quantum systems \citep[][]{rotter_dynamics_2001}. 
\citet{Labbe2015} chose not to show the results, obtained in the spherical case, for values of Le corresponding to avoided crossing (their Figures 6, 7, 11).

We differentiate in the following the modes, characterised by their physical properties, and the eigenbranches, obtained by continuous tracking of the eigensolutions. 
This way, we can continue the $U_3$-mode and the TM out of the $\Le \ll 1$ domain, where they are clearly distinguishable. We have indicated the $U_3$ mode and TM as well as the fastest slow mode by the coloured lines in the bottom Figure \ref{fig:avoidedcrossing}.

\subsubsection{Ellipticity effects}\label{sec:frequencyresults}

\begin{figure}
    \centering
    \includegraphics[width=\linewidth]{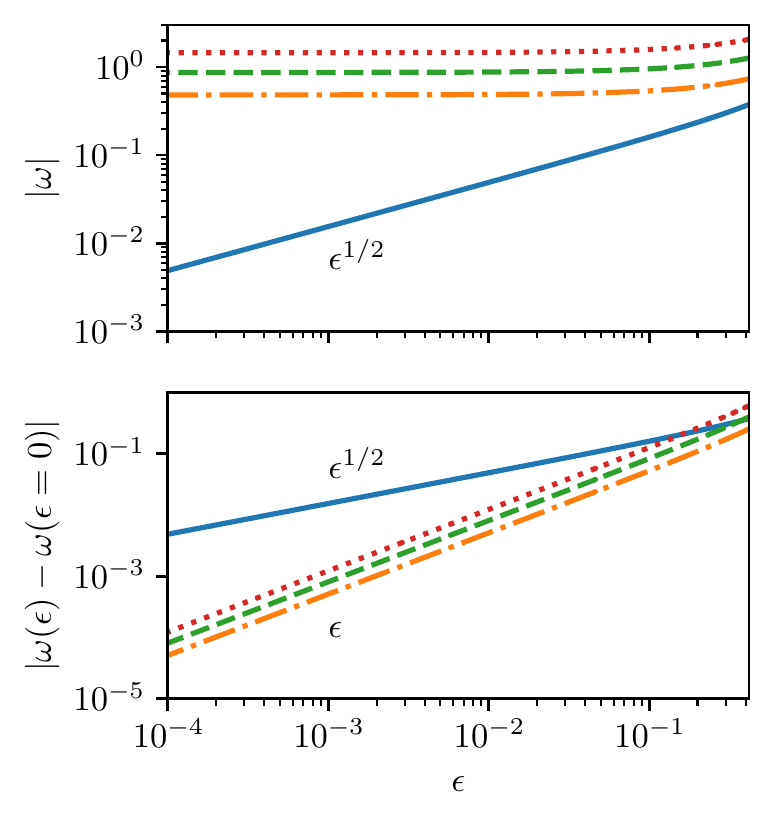}
    \caption{Frequencies of TM (orange dash-dotted, green dashed and red dotted) and the $U_3$-mode (blue solid) as a function of ellipticity $\epsilon$ (top). Difference between the frequency as a function of $\epsilon$ and the frequency in the sphere with $\epsilon=0$ (bottom). The Lehnert number is $\Le=10^{-5}$.}
    \label{fig:freqdeps}
\end{figure}

The dependency of the frequency on the equatorial ellipticity of TM and $U_3$-mode is presented in Figure \ref{fig:freqdeps} (top).
It is observed that, when $\epsilon\lesssim 10^{-1}$, the change of the TM frequency is small and tends to their non-vanishing frequency in the sphere. To be more quantitative, the difference between the frequencies in the ellipsoid and the sphere scales with $\epsilon$ for the TM (see Figure \ref{fig:freqdeps}, bottom).

A very different behaviour is observed for the frequency of the $U_3$-mode, since the frequency itself scales with $\epsilon^{1/2}$. 
This means that the  $U_3$-mode has a vanishing frequency when $\epsilon=0$. 
Also, the ratio of kinetic to magnetic energy of the $U_3$-mode scales with ellipticity. 
These two properties clearly differentiate the $U_3$-mode from TM. 
The restoring force for the $U_3$-mode is the pressure force acting on the elliptical boundary. At small ellipticities it is only the magnetic pressure force.

\subsubsection{Torque balance}

The velocity and magnetic field of an eigensolution $(\tilde\vel_k,\tilde\B_k,\omega_k)$ for a given $\B_0$ and $\Le$ are normalised as
\begin{equation}
    \int_\mathcal{V}\tilde\vel_k\boldsymbol{\cdot}\tilde\vel_k\, \dV + \int_\mathcal{V}\tilde\B_k\boldsymbol{\cdot}\tilde\B_k\, \dV = 1,
\end{equation}
such that they have a unit energy in dimensionless units. We can then calculate the angular momentum $\vec{L}_k$ by inserting $\vel_k$ into \eqref{eq:angularmom}, and its time derivative is given by $\mathrm{i}\omega_k \vec{L}_k$.  
The linearised magnetic pressure torque is given by
\begin{equation}
    \torquepm = -\int_{\mathcal{V}}\pos\times\nabla\left(\tilde\B\bcdot\B_0\right)\, \dV,
\end{equation}
and $\vec{\Gamma}_{\mathrm{pm},k}$ follows by inserting $\tilde\B_k$ and $\B_0$. The hydrodynamic pressure torque \eqref{eq:hydropressuretorque} is calculated by reconstructing the pressure gradient, which cannot be done from the velocity field only.
We reconstruct it instead by inserting $\tilde\vel_k$, $\tilde\B_k$ and $\B_0$ in the momentum equation \eqref{eq:linmom}.

\begin{figure}
  \begin{center}
  \includegraphics[width=\imwidth]{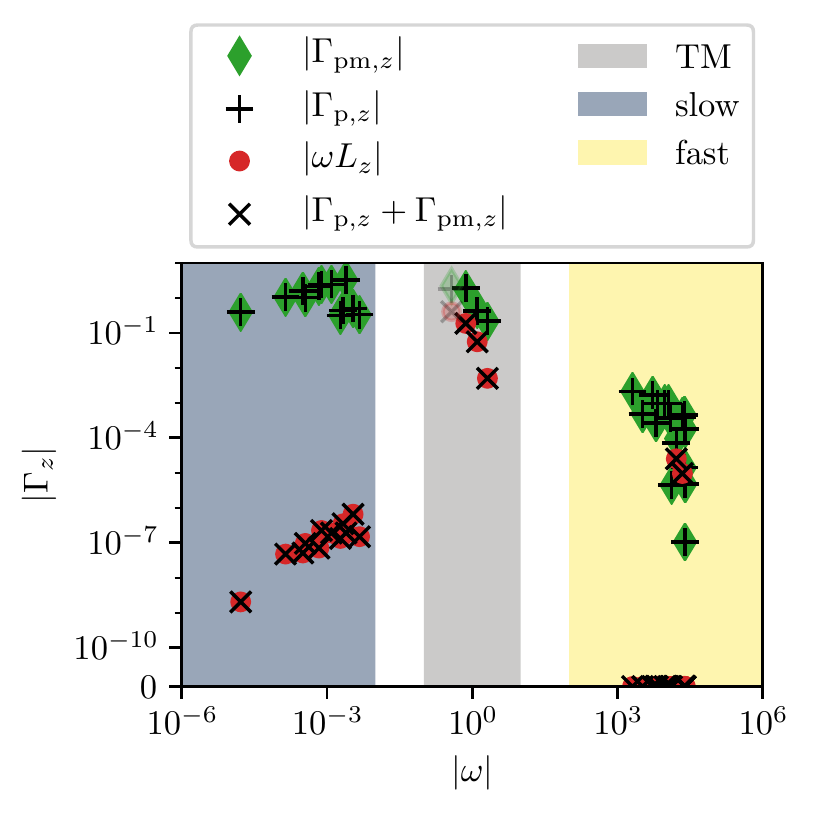}
  \caption{The $z$ component of the torques for the background magnetic field with truncating degree $N=7$, $\epsilon=0.42$ and $\Le=10^{-5}$. The $U_3$-mode is displayed slightly transparent.}
  \label{fig:torques_z}
\end{center}
\end{figure}

The axial torques in a strongly deformed ellipsoid with $a,b,c = 1.25,0.8,1$ (i.e. $\epsilon=0.42$) are shown in Figure \ref{fig:torques_z}. We find non-vanishing torques along the rotation axis for slow modes ($10^{-5}<\omega<10^{-2}$), TM ($10^{-1}<\omega<10$) and the fast modes ($\omega>10^{2}$). 
For many modes the hydrodynamic, magnetic and total pressure (sum of hydrodynamic and magnetic pressure) torques do not vanish. 
For most fast modes $\Gamma_{\mathrm{p},z}$ balances $\Gamma_{\mathrm{pm},z}$ exactly. In case $\Gamma_{\mathrm{p},z}$ is not exactly balanced by $\Gamma_{\mathrm{pm},z}$, the total pressure torque is in balance with the non-vanishing change in angular momentum $\omega L_z$, in agreement with equation \eqref{eq:remainingbalance}. For example, the TM with largest scale (and smallest frequency $\omega=0.737$) has $\Gamma_{\mathrm{p},z}=-1.285+1.414\mathrm{i}$, $\Gamma_{\mathrm{pm},z}=1.159-1.275\mathrm{i}$ and $\mathrm{i}\omega L_z=-0.126+0.139\mathrm{i}$. Our results show that TM yield pressure torques much larger than the slow and fast modes.

\citet{ivers_enumeration_2017} demonstrated that, in the ellipsoid, only flows of uniform vorticity carry angular momentum. 
They are given by
\begin{equation}
\hat\vel_1=\begin{pmatrix}0\\-z/c^2\\y/b^2\end{pmatrix}, \ \hat\vel_2=\begin{pmatrix}-z/c^2\\0\\x/a^2\end{pmatrix}, \ \hat\vel_3=\begin{pmatrix}-y/b^2\\x/a^2\\0\end{pmatrix},
\label{eq:velunivort}
\end{equation}
with a spatially uniform vorticity in the $x$, $y$ and $z$ directions, respectively.
Therefore, we determine if the modes do contain such uniform vorticity components and whether or not it accounts for the non-vanishing angular momentum.
To this end, we must project the eigensolutions onto velocities (\ref{eq:velunivort}) and the resulting angular momentum of the $i$-th uniform vorticity component of an eigensolution with velocity $\vel_k$ is given by
\begin{equation}
    \hat L_{i,k} = \frac{1}{\sqrt{\int\hat\vel_i\boldsymbol{\cdot}\hat\vel_i \, \dV}}\int_\mathcal{V} (\hat\vel_i\boldsymbol{\cdot}\tilde\vel_k)(\hat\vel_i\times\pos)\boldsymbol{\cdot}\vec{1}_i \, \dV.
\end{equation}
For all modes we find (within machine precision) that $L_{z,k}=\hat L_{3,k}$, in agreement with the predictions by \citet{ivers_enumeration_2017}. 

The $U_3$-mode, shown slightly transparent in Figure \ref{fig:torques_z}, has a velocity almost exactly equal to $\hat\vel_3$ (thus the name $U_3$-mode). It is associated with the largest torque. 
However, the time scale at which this torque acts increases as the ellipticity is decreased to more geophysically relevant values, whereas it remains the same for TM.

\subsubsection{Torque variation with the ellipticity and Lehnert number}

\begin{figure}
  \begin{center}
  \includegraphics[width=\imwidth]{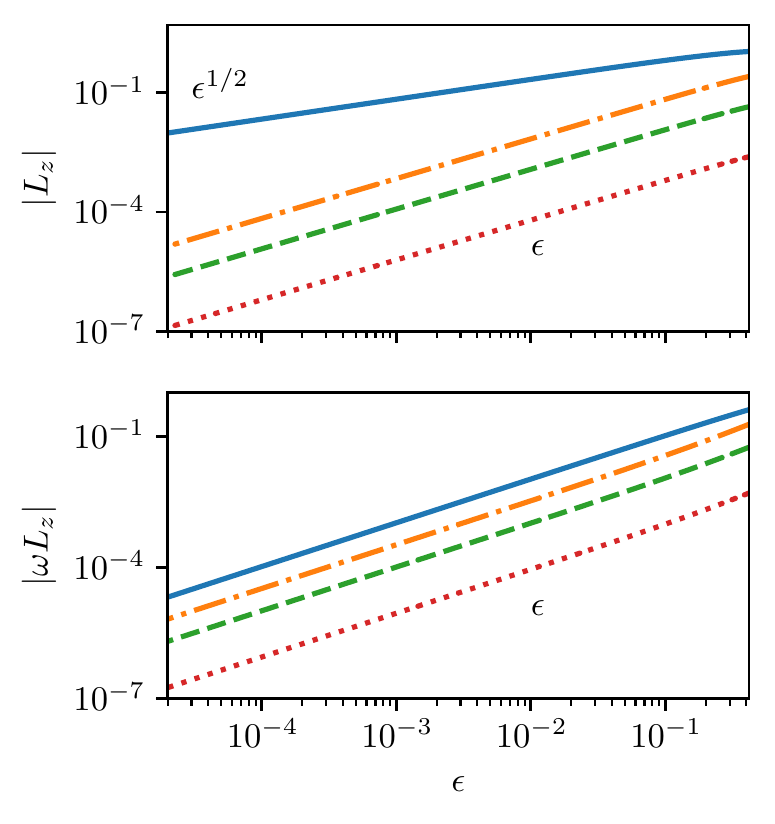}
  \caption{Axial angular momentum (top) and its change (bottom) of TM and the $U_3$-mode as a function of ellipticity for the background magnetic field $\B_{0,\mathrm{QG}}$ and $\Le =10^{-5}$. The colours correspond to those in Figure \ref{fig:freqdeps}. \label{fig:inertialtorque_ellipt_aform_n7_qg}}
\end{center}
\end{figure}

We show in Figure \ref{fig:inertialtorque_ellipt_aform_n7_qg} the dependency on the ellipticity of the angular momentum in $z$ (top), and the associated changes (bottom). The angular momentum scales with $\epsilon$ for the TM, and with $\epsilon^{1/2}$ for the $U_3$-mode. 
Since the frequency is almost independent of $\epsilon$ for the TM, and scales with $\epsilon^{1/2}$ for the $U_3$-mode, the change in angular momentum scales with $\epsilon$ for all modes. 
A vanishing change in angular momentum is necessary to satisfy the torque balance in the sphere, where the pressure torque vanishes exactly. 
Departures from the aforementioned scalings are only observed for strongly deformed ellipsoids (i.e. $\epsilon>0.1$).
For TM with higher frequencies, the spatial complexity of the modes increases and their angular momentum and the change in angular momentum decreases.

\begin{figure}
  \begin{center}
  \includegraphics[width=\imwidth]{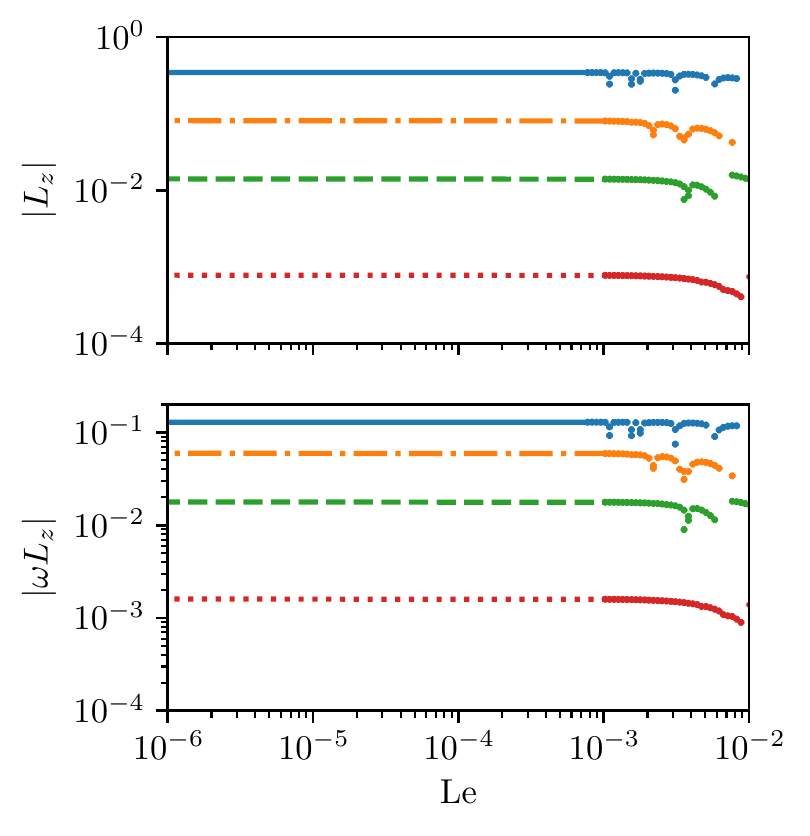}
  \caption{Axial angular momentum (top) and its change (bottom) of TM and the $U_3$-mode for the background magnetic field $\B_{0,\mathrm{QG}}$ and $a,b,c=1.25,0.8,1$ ($\epsilon=0.42$) with $N=7$. At $\Le\gtrsim 10^{-3}$, where eigensolutions are influenced by avoided crossings, we have identified the modes by choosing a frequency within $\pm 10\%$ and an angular momentum within $\pm 50\%$ of the frequency and angular momentum at $\Le\ll 1$. The colours correspond to those in Figure \ref{fig:freqdeps}.} \label{fig:inertialtorque_lehnert_aform_n7_qg}
\end{center}
\end{figure}

In Figure \ref{fig:inertialtorque_lehnert_aform_n7_qg} we show the evolution of the angular momentum and change in angular momentum of TM and the $U_3$-mode as a function of the Lehnert number.
For $\Le\lesssim 10^{-3}$ we observe no dependency on $\Le$ for the angular momentum. 
Thus, because the frequency is also independent of $\Le$ (see Figure \ref{fig:avoidedcrossing}), there is no dependency of the change in angular momentum on $\Le$ and the total pressure torque must scale in the same way.

\begin{figure}
  \begin{center}
  \includegraphics[width=\imwidth]{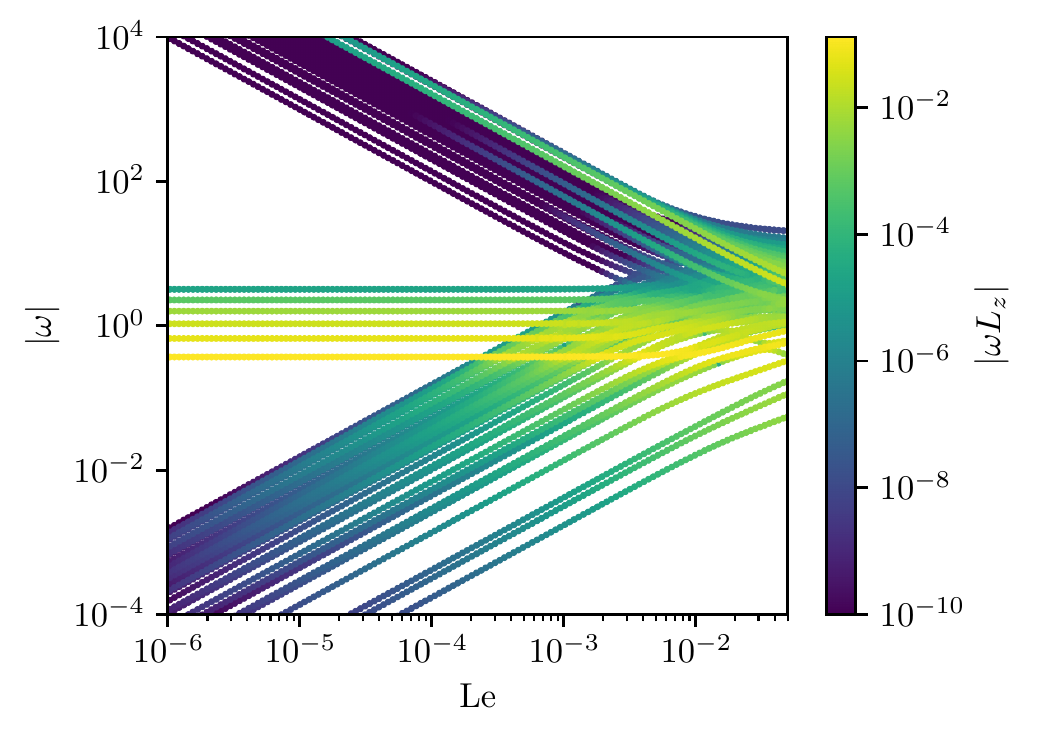}
  \caption{Change of axial angular momentum for the QG model with $\B_{0,\mathrm{QG}}$ and $N=11$. The full spectrum of eigensolutions is computed at incremental steps of $\Le$, without tracking an individual eigensolution.  \label{fig:dense_angularmom}}
\end{center}
\end{figure}

The frequencies of the eigenbranches are close-by when $\Le>10^{-3}$, and they undergo the previously discussed avoided crossings. However, we are still able to identify the TM and $U_3$-mode by their frequency and angular momentum when $\Le\ll 1$ (see Figure \ref{fig:inertialtorque_lehnert_aform_n7_qg}). 
To check the influence of truncation on the results, we have computed the change in angular momentum of all the eigensolutions as a function of $\Le$ for the truncation degree $N=11$.
The $U_3$ mode and the TM can be well characterised by their angular momentum (see Figure \ref{fig:dense_angularmom}). 
Comparison between our results for $N=7$ and $N=11$ makes us confident that our angular momentum calculations for $U_3$ and the largest scale TM are converged at $N=7$.

To check the generality of these results, we have considered a 3-D magnetic field in the hybrid model. The results are presented in Appendix \ref{appendix:hybrid}. 
No qualitative differences with the results of the QG model are found, even if the background magnetic field has a different topology.
We follow from this that our results extend to more complex background magnetic field geometries. 
Further verification was done using a fully 3-D model, where no QG assumption is made on the velocity (not shown).

\section{Discussion and Conclusions}

\subsection{Pressure torque and angular momentum of torsional Alfv\'en modes in ellipsoids}

We have found that TM in the ellipsoid can have a non-vanishing angular momentum. Their angular momentum is fully accounted for by their uniform vorticity flow component along the rotation axis. 
This fully agrees with \citet{ivers_enumeration_2017}, who proved that only uniform vorticity flows have non-zero angular momentum.
In the hydrodynamic case (without magnetic field), only the geostrophic mode can have a non-vanishing axial angular momentum. Since its frequency is zero, the change in axial angular momentum vanishes. 
All other inertial modes in a non-conductive fluid enclosed in an ellipsoid are orthogonal to the geostrophic mode and no inertial mode can produce a net axial torque acting on the boundary. 
In MHD, the modes have a magnetic component and we lose the orthogonality properties between the velocity components.
It is well known that TM exist, whose frequency is non-zero, with a dominant velocity component along the geostrophic contours. 
We have shown that these TM keep following the geostrophic contours for an ellipsoidal domain, and carry angular momentum through their uniform vorticity component when non-axisymmetry is present. 
This change in angular momentum must be balanced by the total pressure torque, as the Coriolis torque is exactly zero along the rotation axis and in our model the magnetic tension torque $\torquebgradb$ also vanishes. 
Our results confirm this balance, and there are modes for which the hydrodynamic pressure torque is larger than the magnetic pressure torque.
It is worth discussing whether our results extend to the case of a perfectly insulating boundary, where the magnetic tension torque $\torquebgradb$ exactly balances the magnetic pressure torque $\torquepm$.
Then, only the hydrodynamic pressure torque can balance changes in angular momentum. 
Investigating perfectly insulating boundaries remains a future problem as it is inherently impossible by our methodology and out of the scope of this work.

We have shown that the frequency of TM remains independent of the Lehnert number, as is the case in the sphere. The angular momentum is also independent of the Lehnert number, and thus is the change of angular momentum and the associated pressure torque. 
The frequency of TM is also almost unaffected by small ellipticities. 
The angular momentum (and its change) of TM scales as $\epsilon$, so that it vanishes in the sphere (as it should).

In addition to the TM, we observed the particular $U_3$-mode, mainly of uniform vorticity in the axial direction, carrying angular momentum.
For a strongly deformed ellipsoid with $\epsilon = \mathcal{O}(10^{-1})$ the frequency of the $U_3$-mode happens to be in the range of TM.
As for TM the frequency and the angular momentum of the $U_3$-mode does not depend on $\Le$ for small enough $\Le$. 
However, in contrast to TM its frequency scales with $\epsilon^{1/2}$, a mode behaviour so far unknown to the authors. The $U_3$-mode is thus geostrophic in the sphere, with a vanishing frequency. Its frequency also vanishes for the hydrodynamic case, regardless of the ellipticity. 
A magnetic field with a component perpendicular to the geostrophic contours is needed in addition to non-axisymmetry to drive this mode.

Another interesting application of our model is the extension to more complex geometries (as long as closed geostrophic contours exist).
The derived equations are indeed independent of the (possibly non-orthogonal) coordinate system. The ellipsoidal case presented here can be used as a benchmark for follow up work in this direction. 
We have additionally presented the first hybrid model, with the velocity in the QG assumption and a 3-D magnetic field. A property highly desirable in core flow dynamics, where a columnar flow model seems appropriate, but the magnetic field is clearly three dimensional \citep[e.g.][]{schaeffer_turbulent_2017}.

\subsection{Geophysical implications}

Our results suggest that TM in the Earth's core, which have periods on the scale of a few years, exert a pressure torque onto the solid mantle, provided the CMB is non-axisymmetric. 
The observed variations in the LOD are $\mathcal{O}(10^{-4})$ s at the 6 yr period \citep{gillet_planetary_2015}, which corresponds to a change in angular momentum $\mathcal{O}(10^{16})$ Nm. 
To compare this to the torques of TM calculated here, we redimensionalise our numerical results by assuming a characteristic velocity $u_0=5\times 10^{-6}$ m/s of TM \citep[see Figure 10 in][]{gillet_planetary_2015}. We match the frequencies of the calculated TM to the 6 yr period, so that a characteristic background magnetic field strength $B_0$ and similarly $\Le$ is defined. 
For an ellipticity $\epsilon=10^{-3}$, estimated for Earth \citep{sze_core_2003,koper_constraints_2003}, the resulting values are presented in Table \ref{tab:estimate_torque}. 
The frequency conversion to match a 6 yr period yields a characteristic magnetic field strength of $B_0\sim 4-30$ mT, hence a Lehnert number $\Le=\mathcal{O}(10^{-4}$). These values are in agreement with what is expected for the Earth's outer core \citep{gillet_fast_2010}. 
The resulting change in angular momentum, and thus the pressure torque, is at most $\mathcal{O}(10^{14})$ Nm for all modes. These values are two orders below the value needed to explain the variation of the LOD on the 6 yr period.
This result can be better understood from a dimensional analysis of the pressure. First, the pressure varies linearly with the TM velocity. 
Second, the TM are independent of $\Omega$. Therefore, the pressure associated to the velocity of TM scales with $p_0\sim \rho u_0 u_A = \mathcal{O}(10^{-3})$ Pa. With this value, we verify that the resulting hydrodynamic pressure torque is $\mathcal{O}(10^{14})$ Nm.

In order to make the pressure torque significant, we need deviations from geostrophy so that pressure depends on $\Omega$. This may happen in the presence of non-closed geostrophic contours. Then, 'pseudo-geostrophic' modes are replaced by Rossby modes, whose properties depend on $\Omega$.
These Rossby modes are not steady and possess the mean circulation included in the geostrophic mode otherwise \citep{Greenspan1968}. 
Thus, Rossby modes driven by the magnetic field may play an important role for the pressure torque on a non-spherical boundary, where non-closed contours exist. 
It is easy to imagine this scenario in the presence of an inner core or at the CMB of the core, with a trough directed inwards at the equator.
Stratification at the upper outer core may further increase the efficiency of the topographic torque \citep{braginsky_magnetic_1998, glane_enhanced_2018,jault_tangential_2020}.

Another hypothetical geophysical application is the explanation of the very long period variations in the LOD through the $U_3$-mode. 
These variations are $\mathcal{O}(10^{-3})$ s and have a period of around 1500 yr \citep{stephenson_long-term_1995,dumberry_azimuthal_2006}.
The $U_3$-mode in our model has a period of 1800 yr for $\Le=10^{-4}$ and an ellipticity of $\epsilon=10^{-3}$. The $U_3$-mode could therefore be an explanation for these long period variations, but this remains a very speculative idea.

\begin{table}
    \centering
    \caption{Estimation of change in angular momentum of TM for Earth's core, with $\epsilon=10^{-3}$. The TM are normalised to have a period $T=6$ yr. The characteristic TM velocity is set to $u_0=5\times 10^{-6}$ m/s.}\label{tab:estimate_torque}
\begin{tabular}{cccccc}
$\mathrm{Model}$ & $\omega$ & $L_z$ & $\mathrm{Le} [10^{-4}]$ & $B_0\, [\mathrm{mT}]$ & $\omega L_z\, [\mathrm{Nm}]$\\
\hline\\
$\mathrm{QG}$ & $0.48$ & $113$ & $9.5$ & $29.5$ & $9.5 \times 10^{14}$\\
$\mathrm{QG}$ & $0.865$ & $18.3$ & $5.28$ & $16.4$ & $1.54 \times 10^{14}$\\
$\mathrm{QG}$ & $1.45$ & $0.92$ & $3.15$ & $9.8$ & $7.76 \times 10^{12}$\\
$\mathrm{Hybrid}$ & $1.14$ & $87.4$ & $4.02$ & $12.5$ & $7.37 \times 10^{14}$\\
$\mathrm{Hybrid}$ & $2.06$ & $9.24$ & $2.22$ & $6.89$ & $7.79 \times 10^{13}$\\
$\mathrm{Hybrid}$ & $3.34$ & $0.35$ & $1.37$ & $4.25$ & $2.95 \times 10^{12}$\\
\end{tabular}
\end{table}

\begin{acknowledgments}
FG was partly funded by Labex OSUG@2020 (ANR10 LABX56).
JN was partly funded by SNF Grant \#200021\_185088.
JV was partly funded by STFC Grant ST/R00059X/1. This work was supported by a grant from the Swiss National Supercomputing Centre (CSCS) under project ID s872.
This work has been carried out with financial support from CNES (Centre National d'\'Etudes Spatiales, France). Support is acknowledged from the European Space Agency through contract 4000127193/19/NL/IA. The authors like to thank two anonymous reviewers for their help in improving this manuscript.
\end{acknowledgments}

\bibliographystyle{gji}
\bibliography{biblio}

\appendix

\section{Geostrophic flow described by a stream function}
\label{appendix:geostrophic}

The geostrophic part of the velocity can be regarded as the average over a geostrophic column. This is equivalent to considering a stream function
\begin{equation}
    \tilde{\psi}(h) =\frac{\oint \psi(h,\varphi) \,  \mathrm{d}\varphi}{\oint \mathrm{d}\varphi},\label{eq:geostream}
\end{equation}
depending only on the geostrophic column height $h$. Here, we have chosen the coordinates $(h,\varphi,z)$ conveniently, such that $\varphi$ is the coordinated along a closed geostrophic contour of constant $h$ and $z\in[-h,h]$ is along the rotation axis. In the axisymmetric case these coordinates are identical to the cylindrical coordinates. 
In the generic case, with arbitrarily shaped geostrophic contours, we have to apply curvilinear coordinates, that are not necessarily orthogonal. For non-orthogonal coordinates the dual, covariant and contravariant, bases $\vec{g}_i$ and $\vec{g}^i$ are needed. We refer the reader to \citet{Aris1989} for more details on non-orthogonal curvilinear coordinates.

Inserting \eqref{eq:geostream} into \eqref{eq:qgvel} the geostrophic velocity is given by
\begin{subequations}
 \label{eq:stream2}
\begin{align}
    \vel_G &= \nabla\tilde{\psi}(h)\times\nabla\left(\frac{z}{h}\right)\\
    &= u_G(h,\varphi)\vec{g}_2,
\end{align}
\end{subequations}
with $u_G(h,\varphi)=(Jh)^{-1}\frac{\partial\tilde\psi}{\partial h}$ and the covariant basis vector in $\varphi$-direction $\vec{g}_2$. 
Here, $J(h,\varphi,z)=\det(g_{ij})$ is the Jacobian of the coordinate mapping. The metric elements are given as $g_{ij}=\vec{g}_i\boldsymbol{\cdot}\vec{g}_j$. In case of the sphere or the ellipsoid $J=J(h)$.
The geostrophic pressure $p_G$ is well defined and depends on $h$ only
\begin{subequations}
\begin{align}
    2\rho \vel_G\times\vec{\Omega} &= -\nabla p_G\\
    \Leftrightarrow \frac{2\rho\Omega}{h}\frac{\partial\tilde{\psi}}{\partial h} &= \frac{\partial p_G}{\partial h}.
\end{align}
\end{subequations}

To construct a basis of geostrophic velocities $\vel_{G,i}$ being polynomial in the Cartesian coordinates the stream function $\psi_i(h)$ has to take the form

\begin{equation}
    \tilde\psi_i(h)=\frac{1}{3}h^{3+2i},\label{eq:geobasis}
\end{equation}
where $h^2=c^2(1-x^2/a^2-y^2/b^2)$. The basis of geostrophic velocities is given as

\begin{equation}
    \vel_{G,i}=\frac{1}{h}\nabla\tilde\psi_i\times\ez=\frac{1}{3}(3+2i)h^{2i}\nabla g\times\ez,
\end{equation}
with $\nabla g=-c^2(x/a^2,y/b^2,0)^\mathsf{T}$.

\section{Hybrid model}\label{appendix:hybrid}

In the hybrid (or fully 3-D) model the background magnetic fields are less restricted, and we select an admissible field from appendix A in \citet{wu_high_2011}. Namely, we consider the magnetic field 

\begin{equation}
    \B_{0,\mathrm{hyb}} = \begin{pmatrix}xy\\-2b^2(x^2/a^2+z^2/c^2)+b^2-y^2\\yz\end{pmatrix},
\end{equation}
named $\vec{v}_8$ in the quadratic basis of \citet{wu_high_2011}. We choose this field, as it clearly goes beyond the magnetic field \eqref{eq:aformb} while keeping the maximum polynomial degree sufficiently low to ensure convergence. 

\begin{figure}
    \centering
      \includegraphics[width=0.49\imwidth]{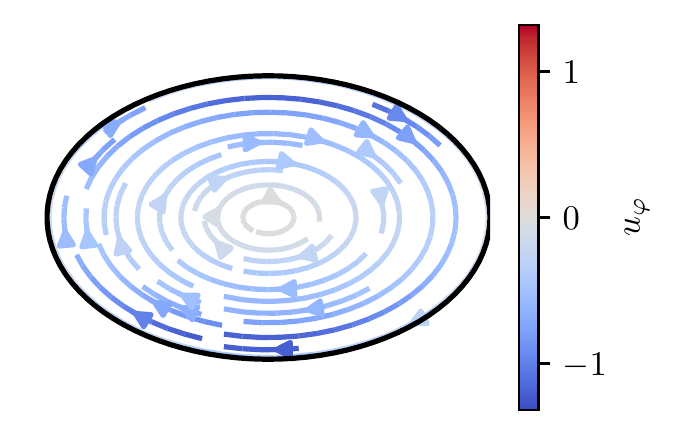}
    \includegraphics[width=0.49\imwidth]{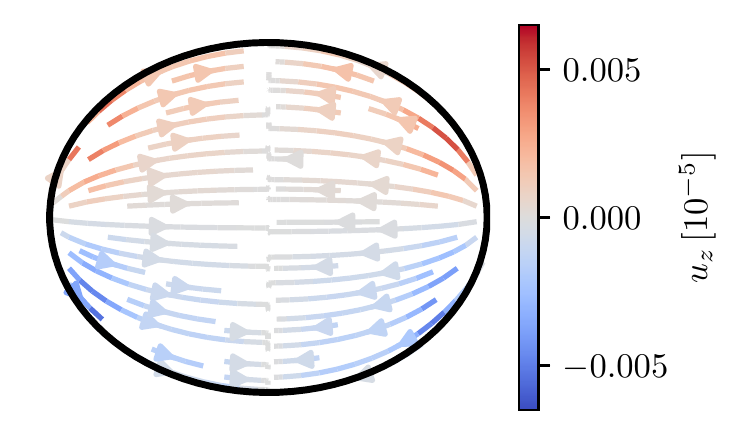}
      \includegraphics[width=0.49\imwidth]{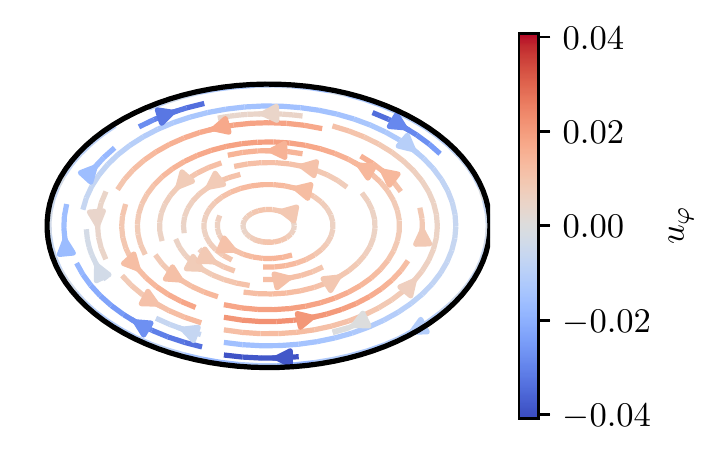}
    \includegraphics[width=0.49\imwidth]{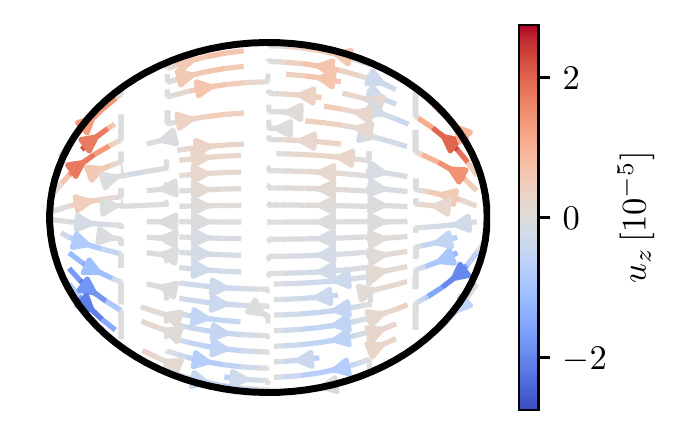}    \includegraphics[width=0.49\imwidth]{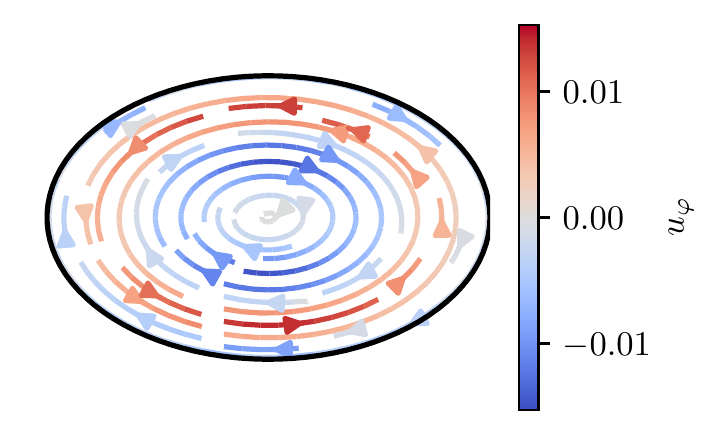}
    \includegraphics[width=0.49\imwidth]{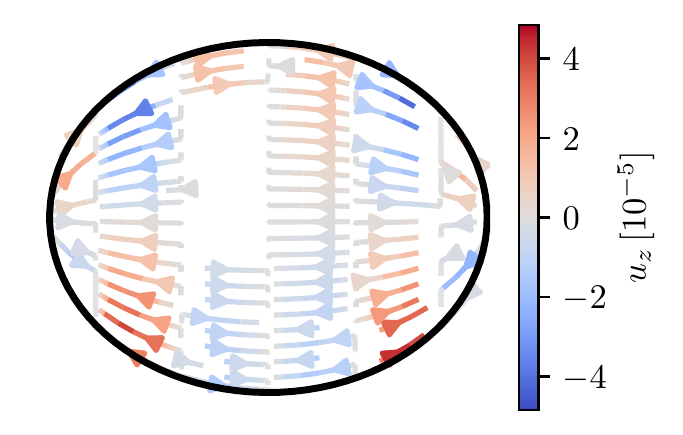}  
    \caption{Equatorial sections (left) and meridional sections along the $x$-axis (right) of the $U_3$-mode (top) and the two largest TM (middle and bottom) using $\B_{0,\mathrm{hyb}}$, $\epsilon=0.42$ and $\Le=10^{-5}$. The colours indicate the velocity along the geostrophic contours $u_\varphi$ and the vertical velocity $u_z$, respectively.}
    \label{fig:streamplots_hyb}
\end{figure}

\begin{figure}
  \begin{center}
  \includegraphics[width=\imwidth]{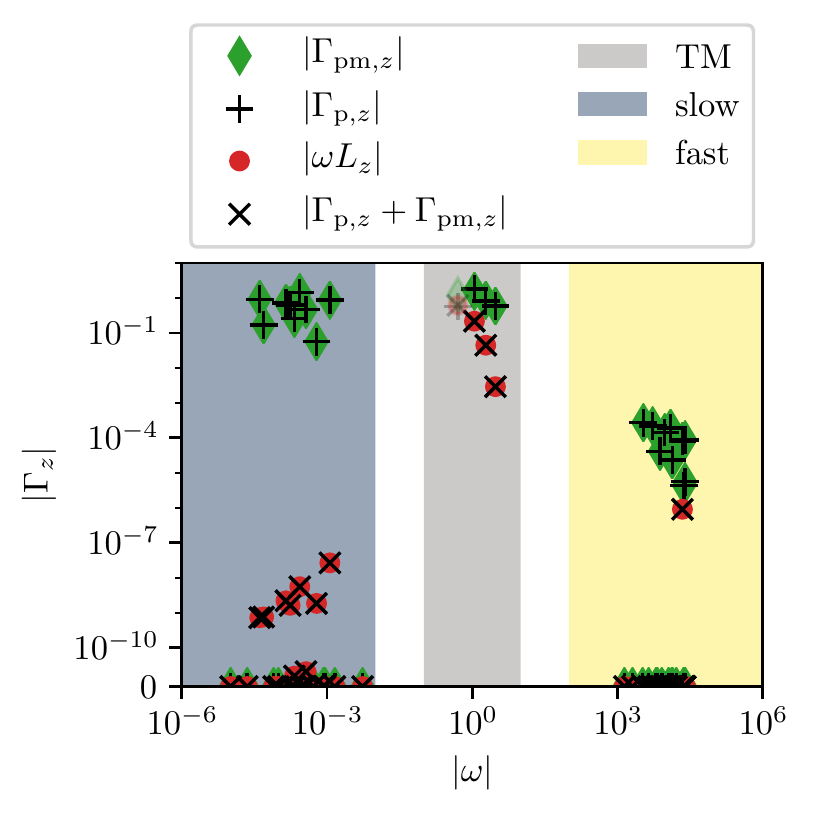}
  \caption{The $z$ component of the torques using $\B_{0,\mathrm{hyb}}$, $\epsilon=0.42$ and $\Le=10^{-5}$ with truncating degree $N=9$.}
  \label{fig:torques_z_hybrid}
\end{center}
\end{figure}

The $U_3$-mode and the two largest scale TM are presented in Figure \ref{fig:streamplots_hyb}. Even though the background magnetic field considered here is topologically speaking very different to $\B_{0,\mathrm{QG}}$, the modes show a clear spatial similarity (compare Figure \ref{fig:streamplots}). 
The axial torques are presented in Figure \ref{fig:torques_z_hybrid}. No qualitative difference to the QG model is observed.
For modes with non-vanishing change in angular momentum the total pressure torque balances it.
Again, the $U_3$-mode carries the largest angular momentum and for some slow modes and fast modes the change in angular momentum is also non-vanishing.

\begin{figure}
  \begin{center}
  \includegraphics[width=\imwidth]{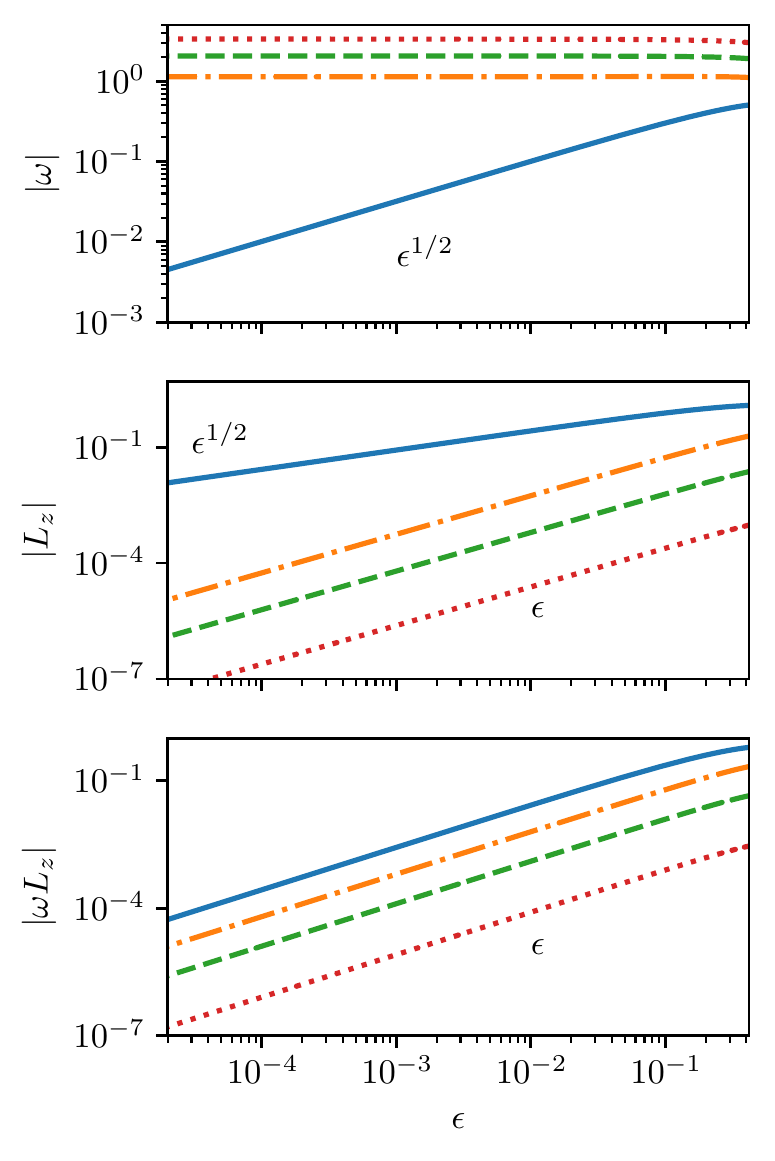}
  \caption{Frequency (top), axial angular momentum (middle) and change in axial angular momentum (bottom) of the three largest scale TM (orange dash-dotted, green dashed and red dotted) and the $U_3$-mode (blue solid) for  $\B_{0,\mathrm{hyb}}$ and $\Le =10^{-5}$ using the hybrid model. \label{fig:inertialtorque_ellipt_b028_n7_hybrid}}
\end{center}
\end{figure}

The dependency of the frequency, angular momentum and the change of angular momentum of the  $U_3$-mode and TM on the ellipticity is shown in Figure \ref{fig:inertialtorque_ellipt_b028_n7_hybrid}. The same scalings in $\epsilon$ are observed for the $U_3$-mode and the TM compared to the QG case.

\begin{figure}
  \begin{center}
  \includegraphics[width=\imwidth]{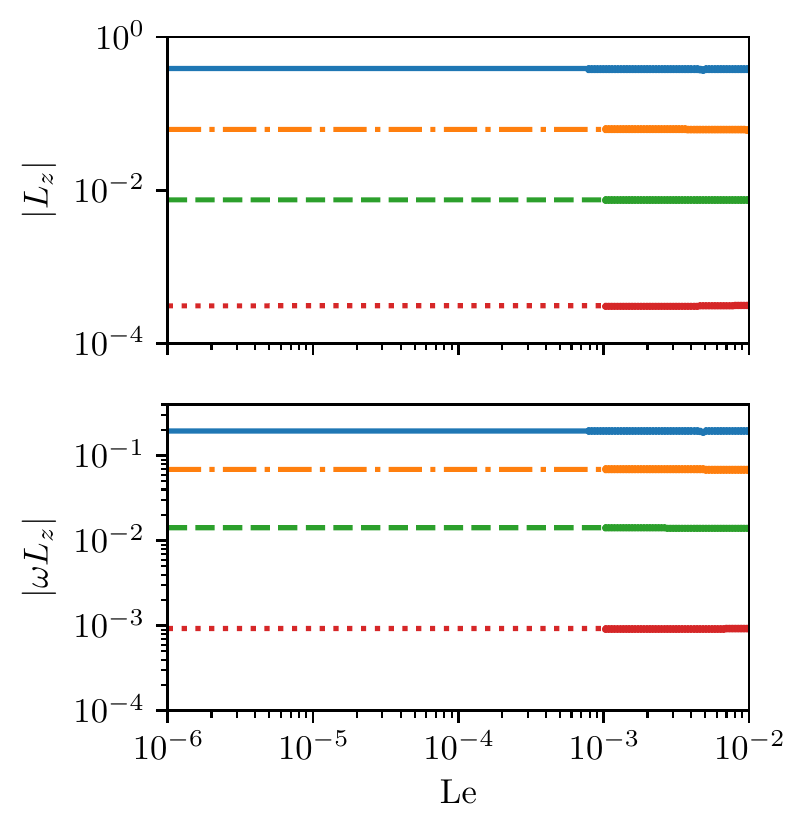}
  \caption{Axial angular momentum (top) and change in axial angular momentum (bottom) for  $\B_{0,\mathrm{hyb}}$ and $a,b,c=1.25,0.8,1$ using the hybrid model. The colours correspond to those in Figure \ref{fig:inertialtorque_ellipt_b028_n7_hybrid}. \label{fig:inertialtorque_lehnert_b028_n7_hybrid}}
\end{center}
\end{figure}

Finally, we present the dependency of the angular momentum and its time derivative of the TM and the $U_3$-mode in Figure \ref{fig:inertialtorque_lehnert_b028_n7_hybrid}. As in the QG case, no dependency is observed. In comparison to the QG case, the $U_3$-mode and the TM seem to be less influenced by avoided crossings at $\Le>10^{-3}$ (compare to Figure \ref{fig:inertialtorque_lehnert_aform_n7_qg}).

\end{document}